\documentclass[journal,draftclsnofoot,onecolumn,12pt]{IEEEtran}
\pdfoutput=1
\usepackage{amsthm,amssymb,graphicx,multirow,amsmath,color,amsfonts}
\usepackage[update,prepend]{epstopdf}
\usepackage[noadjust]{cite}

\usepackage[latin1]{inputenc}
\usepackage{tikz}
\usepackage{bbm} 
\usepackage{pdfpages}
\usepackage{multirow}
\usepackage{comment}

\def\nb0{{\mathbf{0}}}
\def\nb1{{\mathbf{1}}}

\newtheorem{definition}{Definition}

\newtheorem{theorem}{Theorem}

\newtheorem{cor}{Corollary}

\newtheorem{remark}{Remark}

\allowdisplaybreaks 

\begin{document}
\graphicspath{{./Figures/}}
\title{
Coexistence of RF-powered IoT and a Primary Wireless Network with Secrecy Guard Zones
}
\author{
Mustafa A. Kishk and Harpreet S. Dhillon
\thanks{M. A. Kishk and H. S. Dhillon are with Wireless@VT, Department of ECE, Virginia Tech, Blacksburg, VA. Email: \{mkishk,\ hdhillon\}@vt.edu. The support of the U.S. NSF (Grants CCF-1464293, CNS-1617896, and IIS-1633363) is gratefully acknowledged. This paper was presented in part at the IEEE WCNC 2017, San Francisco, CA~\cite{conf_version}. \hfill Last updated: \today.} 
}

\maketitle

\begin{abstract}
This paper studies the secrecy performance of a wireless network (primary network) overlaid with an ambient RF energy harvesting Internet of Things (IoT) network (secondary network). The nodes in the secondary network are assumed to be solely powered by ambient RF energy harvested from the transmissions of the primary network. We assume that the secondary nodes can eavesdrop on the primary transmissions due to which the primary network uses {\em secrecy guard zones}. The primary transmitter goes silent if any secondary receiver is detected within its guard zone. Using tools from stochastic geometry, we derive the probability of successful connection of the primary network as well as the probability of secure communication. Two conditions must be jointly satisfied in order to ensure successful connection: (i) the signal-to-interference-plus-noise ratio (SINR) at the primary receiver is above a predefined threshold, and (ii) the primary transmitter is not silent. In order to ensure secure communication, the SINR value at each of the secondary nodes should be less than a predefined threshold. Clearly, when more secondary nodes are deployed, more primary transmitters will remain silent for a given guard zone radius, thus impacting the amount of energy harvested by the secondary network. Our results concretely show the existence of an optimal deployment density for the secondary network that maximizes the density of nodes that can harvest an amount of energy above a predefined threshold. Furthermore, we show the dependence of this optimal deployment density on the guard zone radius of the primary network. In addition, we show that the optimal guard zone radius selected by the primary network is a function of the deployment density of the secondary network. The interesting coupling between the performance of the two networks is studied using tools from game theory. In addition, we propose an algorithm that can assist the two networks to converge to Nash equilibrium. The convergence of this algorithm is verified using simulations. Overall, this work is one of the few concrete works that symbiotically merge tools from stochastic geometry and game theory. 
\end{abstract}

\section{Introduction} \label{sec:intro}

Owing to rapid technological advances, wireless communication networks are undergoing unprecedented paradigm shifts. One of the most interesting amongst them is the attempt to make wireless networks virtually self-perpetual in terms of their energy requirements. This is especially gaining importance in internet-of-things (IoT) realm where it may not be economical to charge or replace batteries periodically in billions of devices worldwide~\cite{7842431}. In order to achieve this vision of self-perpetual operation, it is necessary  to provide energy harvesting capability to such networks in addition to reducing energy expenditures through energy-efficient communication policies~\cite{DhiHuaJ2013,DhiHuaJ2014} and energy efficient hardware~\cite{jelicic2012analytic}. While, in principle, we can use any available source of energy to power these networks, ambient RF energy harvesting is considered a preferred option due to its ubiquity~\cite{kamalinejad2015wireless}. Further, due to the need of deploying these ambient RF energy harvesting nodes/devices (referred to as {\em energy receivers} or ERs in the rest of this paper) close to the sources of RF signals, concerns on the secrecy of these RF signals were recently raised in the literature~\cite{7018095,6853867,zhang2016artificial,khandaker2015robust,ng2014robust,xu2014multiuser,banawan2016mimo}. While the RF signal source (referred to as {\em primary transmitter} or PT in this paper) transmits confidential messages to legitimate {\em information receivers} (IRs), the existence of ERs close to the PT may enable them to decode these messages. Consequently, careful study of physical layer security in such systems is required to glean insights on the new secrecy performance limitations in the presence of ERs. The importance of physical layer security lies in ensuring the ability of legitimate receivers to decode the confidential messages while preventing illegitimate receivers from decoding these messages~\cite{6772207}. Many solutions were proposed to enhance physical layer security including: (i) using protected zones in order to ensure an eavesdropper-free regions around the transmitters~\cite{romero2014physical,7247765}, (ii) using artificial noise in order to degrade the confidential signal's SNR at the energy receiver~\cite{zhou2010secure,zhang2016artificial}, (iii) beamforming in multi-antenna systems~\cite{khandaker2015robust,ng2014robust,xu2014multiuser}, and (iv) using guard zones to stop information transmission whenever an eavesdropper is detected within a specific region around the transmitter~\cite{5934342,5995290}. 

In this paper, we limit our attention to single-antenna transmitters and receivers, which means beamforming is not applicable to our setup. In addition, using protected zones assumes some physical control over the eavesdroppers, which will not be considered in this paper. Thus, we have two options to choose from: either use artificial noise or guard zones. In~\cite{WCL_1}, it was recently shown that the guard zone technique outperforms the artificial noise technique in the noise limited regime when the link distance is higher than a specific threshold. No such performance comparison between the two techniques is known when interference is taken into account. Regardless, we focus on the guard zone technique in this paper while leaving the artificial noise technique as a promising direction of future work. In particular, modeling the interaction between a primary network using guard zones to enhance secrecy and an IoT network using RF signals transmitted by the primary network to harvest energy is the main focus of this paper.

\subsection{Related Work}
In this subsection, we will provide a brief summary of the related works in three general directions of interest to this paper: (i) stochastic geometry-based analysis of secrecy and/or energy harvesting wireless networks, (ii) game theory-based analysis of secrecy and/or energy harvesting wireless networks, and (iii) analysis of secure wireless power transmission.

{\em Stochastic geometry-based work}. Stochastic geometry has emerged as a powerful mathematical tool for the modeling and analysis of variety of wireless networks~\cite{ElSHosJ2013,ElSSulJ2017,AndGupJ2016,haenggi2012stochastic}. Out of numerous aspects of wireless networks that have been studied using stochastic geometry, two that are most relevant to this paper are: (i) secrecy~\cite{5934342,geraci2014physical,chen2017secrecy}, and (ii) energy harvesting~\cite{DhiLiJ2014,7482720,7841754,sakr2015analysis,flint2015performance}. We first discuss the related works on secrecy. The work presented in~\cite{5934342} quantified the loss in network throughput that results from ensuring a specific level of secrecy performance. In addition, possible performance enhancement using guard zones was also studied. Authors in~\cite{geraci2014physical} studied the physical layer security of downlink cellular networks assuming that all the existing users in the network are potential cooperating eavesdroppers. In~\cite{chen2017secrecy}, authors studied the secrecy of downlink transmission when the transmitter adopts transmit antenna selection. In particular, they derived the secrecy outage probability for the cases of independent and cooperating eavesdroppers, considering both half and full duplex legitimate receivers.  

Stochastic geometry has also been applied to analyze the performance of energy harvesting cellular networks with emphasis on either the downlink channel~\cite{DhiLiJ2014,7482720,7841754} or the uplink channel~\cite{sakr2015analysis}. 
The work presented in~\cite{DhiLiJ2014} provided a comprehensive framework for analyzing heterogeneous cellular networks with energy harvesting base stations (BSs). The primary focus was on characterizing optimal transmission policies as well as quantifying the availability of different tiers of BSs. In~\cite{7482720,7841754}, the joint analysis of the downlink signal quality and the amount of energy harvested at an RF-powered user was performed. In~\cite{sakr2015analysis,flint2015performance}, the uplink counterpart of this problem was explored in which the RF-powered node first harvests energy from ambient RF signals and then uses it to perform data transmission. The work presented in~\cite{7320989} focused on the secrecy analysis of wireless networks where the legitimate transmitters are powered by energy harvesting. To the best of our knowledge, our work provides the first stochastic geometry-based secrecy analysis of wireless power transmission with energy receivers acting as potential eavesdroppers. More details will be provided shortly. 

{\em Game theory-based work}. Tools from game theory have been widely used in the analysis of secrecy in wireless networks~\cite{6654835,6314476,6475179,saad2011distributed}. Using these tools, in~\cite{6654835} authors modeled the interaction between cognitive networks and eavesdroppers, where a channel selection algorithm was proposed to reach the Nash equilibrium. In~\cite{6314476}, authors studied the problem of optimizing the uplink path in multi-hop networks in order to maximize the secrecy rate. In~\cite{6475179}, authors used matching theory to develop an algorithm that enhances the secrecy of source-destination pairs using jamming nodes. In~\cite{saad2011distributed}, authors proposed a distributed algorithm that enhances the achievable secrecy rates in wireless networks with cooperative wireless nodes.

These tools have also been used in the analysis of energy harvesting wireless networks. For instance, authors in~\cite{7815443} provided different approaches to manage energy trading among energy harvesting small cell BSs in order to minimize the consumption of non-renewable energy. Authors in~\cite{6880863} used theses tools to model the relay interference channels where the relay divides the received power from the source into two parts: (i) the first part is used to charge its own battery, and (ii) the rest of the received power is used to forward the received packet to its destination. In~\cite{4300988}, these tools were used to determine optimal probability of switching from listen to active modes and from sleep to active modes for a solar powered wireless sensor network.

{\em Secure wireless power transmission}. While the idea of having the energy receiver as a potential eavesdropper has not been studied yet in the stochastic geometry literature (with randomly located ERs and legitimate transmitters), recent works have explored this idea for the deterministic system setups~\cite{6853867,7018095,7862121,6728676,7879205}. These works assume that the transmitter aims to maximize secrecy performance with the constraint of providing ERs with the required wireless power. In~\cite{6853867,7018095}, authors focused on a single point-to-point link with one ER (potential eavesdropper) in the system. An artificial noise-based solution was proposed to improve secrecy without reducing the amount of wireless power received by the ER. In~\cite{7862121}, authors studied the use of a friendly jammer to enhance the performance of the point-to-point system. The friendly jammer increases the amount of received power by the ER. In addition, it reduces the decodability of the confidential message by the ER. The single link system was extended in~\cite{6728676} to consider one multi-antenna transmitter, one legitimate receiver, and $K$ ERs. Optimal beamforming schemes were provided to either maximize secrecy subject to constraints on harvested energy by the ERs or to maximize harvested energy by the ERs subject to secrecy constraints. Similar approach was adopted in~\cite{7879205} for a more general system model of one macro BS serving $M$ users, $N$ femto BSs each serving $K$ users, and $L$ ERs. Unlike all these works, we will model the locations of ERs and legitimate transmitters using point processes, which will enable us to draw general conclusions that will not be restricted to particular topologies. 

In this paper, we study the secrecy performance of a primary network that consists of PTs and information receivers (which will also be referred to as primary receivers or PRs), overlaid with an IoT network that consists of RF-powered devices (energy receivers or ERs). Guard zones are assumed to be present around PTs in order to improve secrecy. The only sources of ambient RF signals for ERs are the PT transmissions. The PT transmits information (becomes {\em active}) to the PRs only when the guard zone is free of ERs, otherwise it remains silent. On one hand, the IoT network would prefer a dense deployment of ERs so as to increase the overall energy harvested by the IoT network. However, on the other hand, more dense deployment of ERs will mean that more PTs will stay silent (due to the higher likelihood of ERs lying in the guard zones of the PTs) that will ultimately reduce the amount of ambient RF energy harvested, and hence degrade energy harvesting performance. Modeling and analysis of this setup is the main focus of this paper. We summarize the contributions of this paper next.

\subsection{Contributions}
{\em Primary network performance analysis}. Modeling the locations of PTs and ERs by two independent Poisson point processes (PPPs), we show that the locations of {\em active} PTs (PTs with ER-free guard zones) follow Poisson hole process. For this setup, we define the probability of successful connection ($P_{\rm con}$) between the PT and its associated PR by the joint probability of the PT being active and the SINR at the PR being above a predefined threshold. We derive $P_{\rm con}$ as a function of the density of ERs and the guard zone radius $r_g$. We concretely derive a threshold on the density of the PTs below which $P_{\rm con}$ is a decreasing function of $r_g$. Above this threshold, we prove the existence of an optimal value of $r_g$ that maximizes $P_{\rm con}$. For the secrecy analysis, we derive the probability of secure communication (defined by the probability of having the SINR value at any ER less than a predefined threshold). Referring to this metric as $P_{\rm sec}$, we provide several useful insights on the effect of the PT transmission power as well as the density of ERs on the value of $P_{\rm sec}$. 

{\em IoT network performance analysis}. For the IoT network (secondary network), we derive the probability of harvesting a minimum amount of energy $E_{\rm min}$ by the ERs. We define the density of ERs that satisfy this condition as the {\em density of successfully powered ERs}. We prove the existence of an optimal deployment density of ERs that maximizes this density of successfully powered ERs. In order to capture the relation between this optimal density and the guard zone radius, we derive a useful lower bound on the the density of successfully powered ERs. We show that the optimal deployment density that maximizes this lower bound is a decreasing function of $r_g$. Although this conclusion is drawn using a lower bound, we use numerical results and simulations to demonstrate that it holds for the exact expressions as well.

{\em Modeling the interaction between the two networks}. Building on the above results, we show that the interaction between the two networks can be modeled using tools from game theory. In particular, we show that this system can be modeled by a two player non-cooperative static game. The first player is the primary network with the guard zone radius representing its strategy. Its utility function is modeled to capture the successful connection probability as well as the probability of secure communication. The second player is the IoT network with the deployment density representing its strategy. Its utility function is modeled to capture the main performance metric of this network, which is the density of successfully powered ERs. We propose a best response-based learning algorithm that helps in achieving the Nash equilibrium of this game.

\section{System Model} \label{sec:SysMod}
\begin{figure}
\centering
\includegraphics[width=0.6\columnwidth]{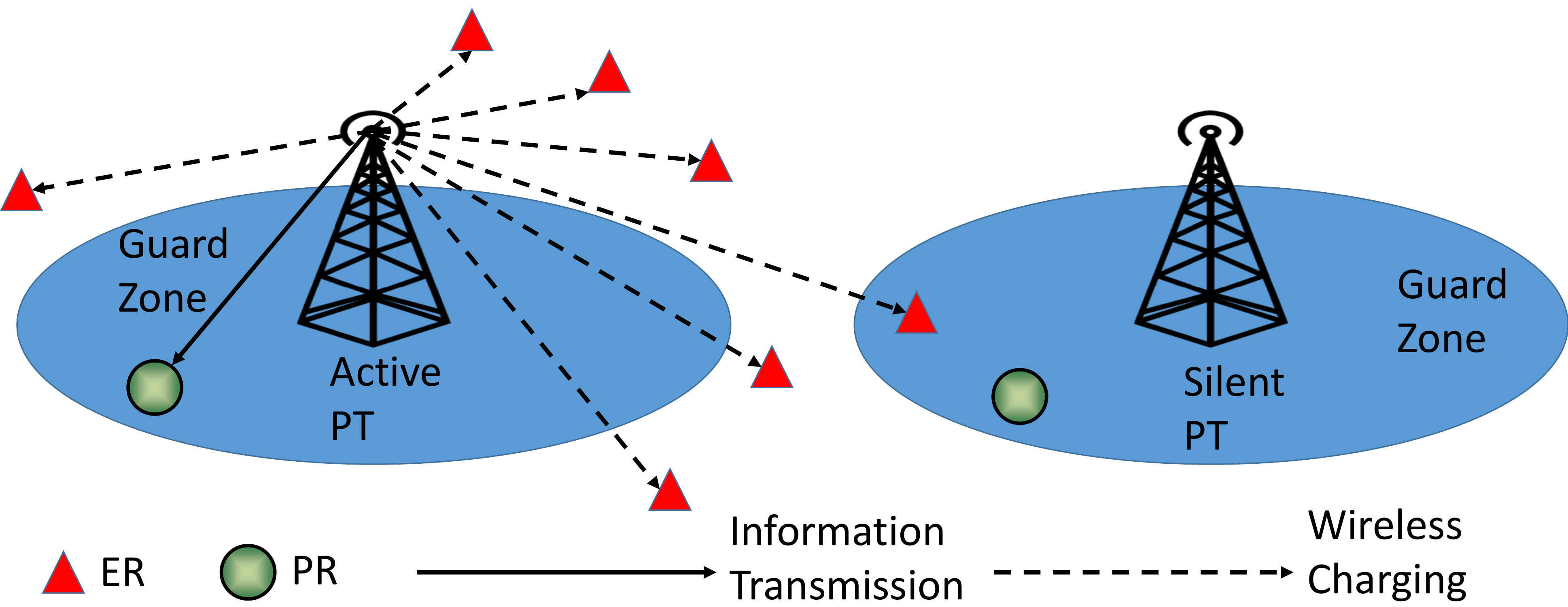}
\caption{The PT stops transmission (becomes silent) if any ER is detected within its guard zone.}
\label{fig:sys1}
\end{figure}

We consider a system that is composed of two wireless networks: (i) a primary network, and (ii) an IoT network (will be referred to as secondary network in the rest of this paper). The primary network is constructed of PTs and primary receivers (will be referred to as either PRs or {\em legitimate receivers} interchangeably throughout the paper). The locations of the PTs are modeled by a homogeneous PPP $\Phi_P\equiv\{x_{i}\}\subset\mathbb{R}^2$ with density $\lambda_P$. In order to enhance secrecy performance, each PT is surrounded by a circular guard zone of radius $r_g$ centered at the PT. We assume that each PT is able to detect the presence of any illegitimate receiver within its guard zone~\cite{6787008}. Various detecting devices can be used for this purpose including metal detectors and leaked local oscillator power detectors~\cite{romero2014physical}. The benefits of using secrecy guard zones and their effect on secrecy performance were discussed in~\cite{5934342}. As shown in Fig.~\ref{fig:sys1}, before the PT transmits data to its associated PR, it scans the guard zone for any illegitimate receivers. If the guard zone is clear, the PT transmits the confidential data, otherwise, the PT stops transmission (becomes {\em silent}). The secondary network is constructed of RF-powered nodes that harvest RF energy from the signals transmitted by the primary network. We refer to these nodes as energy receivers (will be referred to as either ERs or {\em eavesdroppers} interchangeably throughout the paper). The locations of the ERs are modeled by a homogeneous PPP $\Phi_S\equiv\{y_{i}\}\subset\mathbb{R}^2$ with density $\lambda_S$. From the primary network's perspective, the ERs are considered illegitimate receivers. Hence, applying the guard zone scheme, the PT stops transmission whenever at least one ER exists within its guard zone. We assume that ERs are the only potential eavesdroppers in the system. 

We focus our analysis on the typical PT whose intended PR is located at a given distance $r_1$ from the PT. Drawing analogy from the Poisson bipolar model, we call this the {\em typical link} and its constituent PT and PR as {\rm typical PT} and {\rm typical PR}, respectively. Note that this terminology is indeed rigorous if we assume a Poisson bipolar model in which {\em each} PT has an associated PR at a fixed distance. However, since the same setup can be used for cellular networks by treating $r_1$ as the {\em conditional} value of the serving distance, we leave the setup general. Overall, the assumption of fixed $r_1$ enables us to gain several useful insights. In particular, this enables us to better understand how the rest of the system parameters (e.g. $r_g$, $\lambda_S$, and PT's transmission power) affect the interaction between the two networks.  Due to the stationarity of PPP, the typical PR can be assumed to be located at the origin without loss of generality. The typical PT is located at $x_1$ at distance $r_1=\|x_1\|$ from the typical PR. In case the guard zone is clear of ERs, the typical PT transmits the confidential message to the typical PT. In that case, the received power at the typical PR is $P_t h_1r_1^{-\alpha}$, where $P_t$ is the PT transmission power, $h_1\sim\exp(1)$ models Rayleigh fading gain, and $r_1^{-\alpha}$ models power law path-loss with exponent $\alpha > 2$.

\begin{table}[t]\caption{Table of Notations}
\centering
\begin{center}
\resizebox{\textwidth}{!}{
    \begin{tabular}{ {c} | {c} }
    \hline\hline
    \textbf{Notation} & \textbf{Description} \\ \hline
    $\Phi_{P}$; $\lambda_{P}$ & PPP modeling the locations of the PTs; density of the PTs \\ \hline
    $\Phi_{S}$; $\lambda_S$ & PPP modeling the locations of ERs; density of the ERs \\ \hline
        $E_{\rm H}$ & Amount of energy harvested by ER (in Joules)\\ \hline
        $P_{\rm con}$ & The probability of successful connection between a PT and its associated PR\\ \hline
        $P_{\rm sec}$ & The probability of secure communication\\ \hline
        $P_{\rm energy}$ & The density of successfully powered ERs\\ \hline
        $h$; $g$; $w$ & Fading gains for the link between the typical PR and an arbitrary PT; the typical PT and an arbitrary ER; an ER and its nearest active PT\\ \hline
        $r_g$; $R_e$ & Guard zone radius; distance between a PT and its nearest ER\\ \hline
        $\mathbbm{1}(\Xi)$ & An indicator function such that $\mathbbm{1}(\Xi)=1$ if the condition $\Xi$ is satisfied and equals zero otherwise\\ \hline        
        $P_{\rm t}$; $\sigma_P^2$; $\sigma_S^2$ & PT transmission power; noise power at the PR; noise power at the ER\\ \hline
        $\epsilon$; $\alpha$ & Minimum required value for $P_{\rm sec}$; path loss exponent ($\alpha>2$)\\ \hline
        $\beta_P$; $\beta_S$ & Minimum required SINR value at the PR to ensure successful connection; threshold on SINR at ERs to ensure perfect secrecy\\ \hline\hline
    \end{tabular}}
\end{center}
\label{tab:TableOfNotations}
\end{table}
\subsection{Primary Network Modeling}
According to Wyner's encoding scheme~\cite{6772207,4276938} and the approach used in~\cite{5934342}, the PT defines the rate of codewords and the rate of confidential messages, $C_{\rm t}$ and $C_{\rm s}$ respectively. The difference, $C_{\rm e}=C_{\rm t}-C_{\rm s}$, can be interpreted as the cost paid to secure the confidential messages. Let the mutual information between PT's channel input and PR's channel output be $\mathcal{I}_{\rm t}$ and between PT's channel input and ER's channel output be $\mathcal{I}_{\rm e}$. Then, the objective is to ensure that the following two conditions are satisfied: (i) $C_{\rm t}\leq\mathcal{I}_{\rm t}$ to ensure successful decoding at the PR, and (ii) $C_{\rm e}\geq\mathcal{I}_{\rm e}$ to ensure perfect secrecy. Equivalently, we can define two SINR thresholds at the legitimate receiver and at the eavesdropper. To ensure successful decoding of the received confidential message at the PR, the condition ${\rm SINR_P}\geq\beta_P$ should be satisfied, where $\beta_P=2^{C_{\rm t}}-1$. Similarly, to ensure perfect secrecy, the SINR at any eavesdropper should satisfy the condition ${\rm SINR_S}\geq\beta_S$, where $\beta_S=2^{C_{\rm t}-C_{\rm s}}-1$.
\begin{figure}
\centering
\includegraphics[width=0.8\columnwidth]{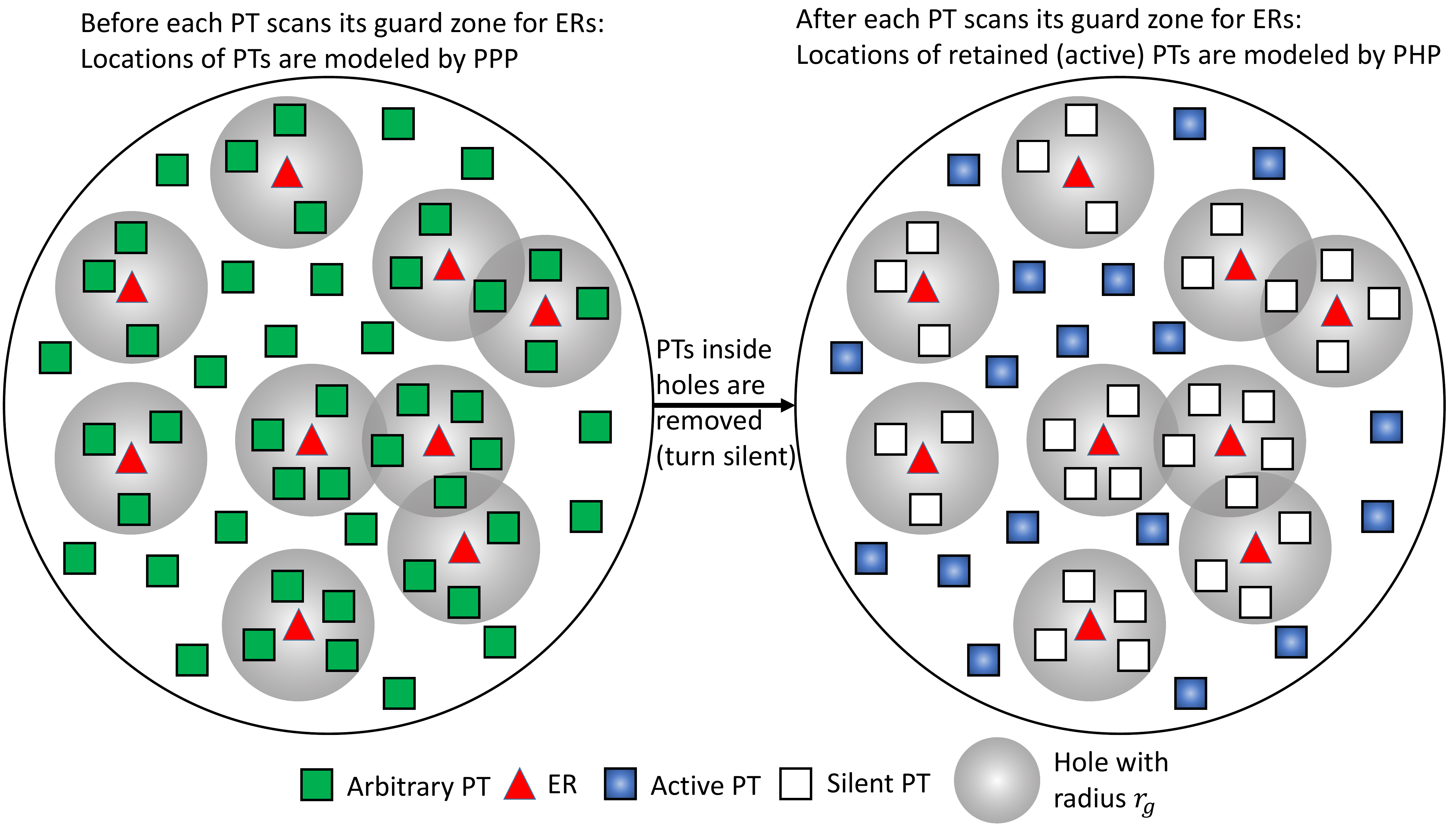}
\caption{The locations of active PTs is modeled by a PHP where the locations of ERs represent the centers of the holes.}
\label{fig:sys2}
\end{figure}
In order to mathematically define the instantaneous value of SINR at the legitimate receiver, we need to capture the effect of adopting the guard zone scheme on the interference level at the typical PR. According to the guard zone scheme, an interfering PT located at $x_i$ is active only if its guard zone is clear of ERs. Hence, denoting the random variable representing the distance between the typical PT and its nearest eavesdropper by $R_e$, the probability of a typical PT being active is
\begin{align}
\label{1}
P_{\rm active}=\mathbb{P}(R_e\geq r_g)=e^{-{\lambda}_{S}\pi{r_g}^2}.
\end{align}
Next, we formally define the value of SINR at the typical PR as
\begin{align}
\label{2}
{\rm SINR_P}=\frac{P_th_1r_1^{-\alpha}}{\sum_{x_i\in\Phi_P\backslash{x_1}}\delta_iP_th_i\|x_i\|^{-\alpha}+\sigma_P^2},
\end{align}
where $h_i\sim \exp(1)$ models Rayleigh fading gain for the link between the typical PR and the PT located at $x_i$, and $\sigma_P^2$ is the noise power at the PRs. The indicator function $\delta_i$ is used to capture only the {\em active} interfering nodes. Hence, $\delta_i=1$ if the PT located at $x_i$ is active and equals zero otherwise. The expected value of $\delta_i$ is $\mathbb{E}[\delta_i]=P_{\rm acitve}$.

The locations of the active PTs can be modeled using Poisson hole process (PHP) $\Psi$~\cite{YazDhiJ2016,6155562}. A PHP is constructed using two independent PPPs: (i) Baseline PPP $\Phi_1$, and (ii) the PPP modeling the locations of the hole centers $\Phi_2$. All points of $\Phi_1$ that are within distance $D$ from any point in $\Phi_2$ are carved out. The remaining points of $\Phi_1$ represent the PHP. In our system, as shown in Fig.~\ref{fig:sys2}, we can use the same approach to model the locations of the active PTs. Any PT in $\Phi_P$ that is within a distance $r_g$ from any ER in $\Phi_S$ is inactive. Hence, the locations of the remaining (active) PTs are modeled by PHP. This can be formally defined as follows
\begin{align}
\label{3}
\Psi=\left\{x\in\Phi_P:x\notin\bigcup_{y\in\Phi_S}\mathcal{B}(y,r_g)\right\},
\end{align}
where $\mathcal{B}(y,r_g)$ is a ball of radius $r_g$ centered at $y$. Since the probability of retention of any point in $\Phi_P$ is $P_{\rm active}$, the density of the resulting PHP $\Psi$ is $\lambda_PP_{\rm acitve}$. 

The primary network tries to jointly optimize two main performance metrics: (i) successful connection probability, and (ii) secure communication probability. While the former is related to the successful delivery of the data from the PT to the PR, the latter is related to the security of the transmitted data when the PT is active. Both metrics are formally defined below. 
\begin{definition}[Probability of successful connection] \label{def:1}
In order to ensure successful connection between the typical PT and PR, two conditions need to be satisfied: (i) the typical PT is active, and (ii) the SINR at the typical PR is greater than the threshold $\beta_P$. Therefore, the probability of successful connection is
\begin{align}
\label{4}
P_{\rm con}(r_g,\lambda_S)=\mathbb{P}(R_e\geq r_g,{\rm SINR_P}\geq\beta_P).
\end{align}
\end{definition}

When a PT is transmitting confidential data (active PT), the condition of ${\rm SINR_S}\leq\beta_S$ needs to be satisfied at each ER in order to ensure perfect secrecy. Focusing on the signal transmitted by the typical PT, the SINR value at an ER located at $y_j$ is 
\begin{align}
\label{5}
{\rm SINR_S}(y_j)=\frac{P_tg_{1,j}\|x_1-y_j\|^{-\alpha}}{\sum_{x_i\in\Phi_P\backslash{x_1}}\delta_iP_tg_{i,j}\|x_i-y_j\|^{-\alpha}+\sigma_S^2},
\end{align} 
where $g_{i,j}\sim \exp(1)$ models Rayleigh fading gain for the link between the PT located at $x_i$ and the ER located at $y_j$, and $\sigma_S^2$ is the noise power at the ERs. We now define the second performance metric for the primary network, the secure communication probability, next.
\begin{definition}[Secure communication probability] \label{def:2}
Given that a PT is active, the probability that its transmitted data is perfectly secured is
\begin{align}
\label{6}
P_{\rm sec}(r_g,\lambda_S)=\mathbb{E}\left[\mathbbm{1}\left(\bigcup_{{y_j}\in\Phi_S}{\rm SINR_S}(y_j)\leq\beta_S\Bigg|R_e\geq r_g\right)\right].
\end{align}
\end{definition}
For a given value of $\lambda_S$, the primary network selects the value of $r_g=r_g^*$ that maximizes the value of $P_{\rm con}$ while ensuring $P_{\rm sec}\geq\epsilon$, where $0<\epsilon\leq 1$. The selection of $r_g^*$ is mathematically formulated next. 
\begin{definition}[Guard zone radius selection] \label{def:3}
The primary network selects the guard zone radius $r_g^*$ that satisfies the following 
\begin{align}
\label{7}
r_g^*&=\arg\underset{r_g\in\mathcal{G}(\lambda_S)}{\max}\ P_{\rm con}(r_g,\lambda_S),\nonumber\\
\mathcal{G}(\lambda_S)&=\{r_g:P_{\rm sec}(r_g,\lambda_S)\geq\epsilon\}.
\end{align} 
\end{definition}
\begin{remark}
\label{teaser}
By observing Eq.~\ref{7}, we note that the value of $r_g^*$ selected by the primary network is a function of the density of the secondary network $\lambda_S$. In addition, while $P_{\rm active}$ is obviously a decreasing function of $r_g$, the exact effect of $r_g$ on either $P_{\rm con}$ or $P_{\rm sec}$ is harder to describe than one expects. It might seem intuitive to expect that as $r_g$ increases the average distance between any PT and its nearest ER will increase, leading to an increase in the probability of secure communication. However, this is actually not always correct. The reason for that is the effect of $r_g$ on the density of the PHP $\Psi$, $\lambda_PP_{\rm acitve}$, which is the density of {\em active} interferers. As the guard zone radius increases, the density of interferers decreases, which means that the interference level at any ER will decrease on average leading to a {\em larger} SINR value.
\end{remark}
To better understand the behavior of all the aforementioned performance metrics, we will derive some insightful expressions that will, with the help of the numerical results, provide a complete picture for the effect of the system parameters on these metrics. 
\subsection{Secondary Network Modeling}
For an ambient RF energy harvesting device, the distance to the nearest source is critical and has major effect on the average harvested energy value, as shown in~\cite{7841754}. The energy harvested from the nearest source is frequently used in the literature to study the performance of ambient RF energy harvesting wireless devices. For instance, in~\cite{6575083}, authors proposed the concept of harvesting zone where an ER is able to activate its power conversion circuit and harvest energy only if it is within a specific distance from the active PT. Consequently, we focus our analysis of the energy harvested from the nearest PT since it is the dominant source of ambient RF energy. Hence, the energy harvested by the ER located at $y_j$ is
\begin{align}
\label{8}
E_{H}=\eta T P_t\|x_{j,1}-y_j\|^{-\alpha}w_j,
\end{align}
where $\eta$ models the RF-DC efficiency of the ER, $T$ is the duration of the transmission slot (assumed to be unity in the rest of the paper), $x_{j,1}$ is the location of the nearest active PT to the ER located at $y_j$, and $w_j\sim \exp(1)$ models Rayleigh fading gain for the link between the ER located at $y_j$ and its nearest active PT. Most of the relevant existing works use one of two performance metrics for the anlysis of energy harvesting wireless networks: (i) the expected value of the harvested energy $\mathbb{E}[E_H]$ as in~\cite{flint2015performance}, or (ii) the energy coverage probability $\mathbb{P}(E_H\geq E_{\rm min})$ as in~\cite{sakr2015analysis}, where $E_{\rm min}$ is the minimum required value of $E_H$ at each ER. We will use a modified version of the latter metric. In order to maximize the density of ERs that harvest the minimum amount of energy $E_{\rm min}$, the secondary network selects the network density $\lambda_S=\lambda_S^*$ that maximizes 
\begin{align}
\label{9}
P_{\rm energy}=\lambda_S\mathbb{P}(E_H\geq E_{\rm min}).
\end{align}
The above metric represents the {\em density of the successfully powered ERs}.
\begin{remark}
\label{teaser_2}
Note that the distance between an ER and its nearest active PT increases, on average, as the density of active PTs, $\lambda_PP_{\rm acitve}$, decreases. Recalling, from Eq.~\ref{1}, that $P_{\rm active}$ is a decreasing function of both $r_g$ and $\lambda_S$, we can expect $\mathbb{P}(E_H\geq E_{\rm min})$ to be a decreasing function of both $r_g$ and $\lambda_S$. This implies two things: (i) the performance metric defined in Eq.~\ref{9} has a local maximum at $\lambda_S=\lambda_S^*$, and (ii) the value of the selected density $\lambda_S^*$ that maximizes Eq.~\ref{9} is a function of $r_g$.
\end{remark} 
From Remarks~\ref{teaser} and~\ref{teaser_2} we can get some initial observations on the coupling between the two networks. To better understand the relation between the parameters selected by each of the networks ($r_g$ for the primary network, and $\lambda_S$ for the secondary network), we first need to derive an expression for each of $P_{\rm con}$ and $P_{\rm sec}$ for the primary network, and $P_{\rm energy}$ for the secondary network. 
\section{Performance Analysis}
\subsection{Primary Network}
\subsubsection{Probability of Successful Connection}
As stated earlier, for a given value of $\lambda_S$, the primary network selects the optimal guard zone radius $r_g^*$ that maximizes $P_{\rm con}$ while ensuring that $P_{\rm sec}\geq\epsilon$. The process of $r_g$ selection was mathematically formulated in Definition~\ref{def:3}. In the following Theorem we derive $P_{\rm con}$.
\begin{theorem}[Probability of successful connection]
\label{thm:Pcon1}
For a given value of $\lambda_S$, the probability of successful connection defined in Definition~\ref{def:1} is
\begin{align}
\label{10}
P_{\rm con}(r_g,\lambda_S)=\exp\left(-\left[\lambda_S\pi r_g^2+\beta_P\frac{\sigma_P^2}{P_t}r_1^{\alpha}+\frac{2\pi^2\lambda_P P_{\rm active}\beta_P^{\frac{2}{\alpha}}r_1^{2}}{\alpha\sin(\frac{2}{\alpha}\pi)}\right]\right).
\end{align}
\end{theorem}
\begin{IEEEproof}
See Appendix~\ref{app:Pcon}.
\end{IEEEproof}
\begin{remark}
\label{rem:Pcon}
According to Definition~\ref{def:1} of $P_{\rm con}$, the effect of $r_g$ on $P_{\rm con}$ is two fold. On one hand, increasing the value of $r_g$ decreases the probability of the PT being active, which decreases $P_{\rm con}$. This effect appears in the first term in the exponent in Eq.~\ref{10}. On the other hand, increasing the value of $r_g$ decreases the density of active interferers. Hence, the probability of having SINR higher than $\beta_P$ increases, which increases $P_{\rm con}$. This effect appears in the third term in the exponent in Eq.~\ref{10} implicitly in the value of $P_{\rm active}$.
\end{remark}
To better understand the effect of $r_g$ on $P_{\rm con}$, we derive the value of $\hat{r}_g$ that maximizes $P_{\rm con}$ in the following Theorem.
\begin{theorem}
\label{thm:Pcon2}
Defining $\mathcal{A}_1=\frac{2\pi^2\lambda_P \beta_P^{\frac{2}{\alpha}}r_1^{2}}{\alpha\sin(\frac{2}{\alpha}\pi)}$ then:
\begin{itemize}
\item If $\mathcal{A}_1\leq 1$, then $P_{\rm con}$ is a decreasing function of $r_g$, and $\hat{r}_g=0$.
\item If $\mathcal{A}_1> 1$, then $\hat{r}_g=\sqrt{\frac{\ln\left({\mathcal{A}_1}\right)}{\pi\lambda_S}}$.
\end{itemize}
\end{theorem}
\begin{IEEEproof}
See Appendix~\ref{app:Pcon2}.
\end{IEEEproof}
\begin{remark}
\label{rem:Pcon2}
From the above Theorem, we conclude that when $\mathcal{A}_1\leq 1$ the effect of $r_g$ on the density of active interferers is negligible. This is consistent with intuition. Observing the expression of $\mathcal{A}_1$ in Theorem~\ref{thm:Pcon2}, we note that it is an increasing function of $\lambda_P$, $\beta_P$, and $r_1$. 
Hence, $\mathcal{A}_1\leq 1$ results from one or more of the following conditions: (i) $\lambda_P$ is too small to take the effect of interference into consideration (the system is noise limited), (ii) $r_1$ is too small such that the received signal power from the intended PT is hardly attenuated by the path-loss, and (iii) $\beta_P$ is a very relaxed threshold. These three consequences of having $\mathcal{A}_1\leq 1$ make the condition of $R_e\geq r_g$ in Definition~\ref{def:1} of $P_{\rm con}$ dominate the other condition of ${\rm SINR_P}\geq\beta_P$. This eventually makes $P_{\rm con}$ a decreasing function of $r_g$, similar to $\mathbb{P}(R_e\geq r_g)$.
\end{remark}
In the following corollary, an upper bound on the value of $P_{\rm con}(r_g^*,\lambda_S)$ is provided.
\begin{cor}
Using the result in Theorem~\ref{thm:Pcon2}, the value of $P_{\rm con}(r_g^*,\lambda_S)$ is upper bounded as follows
\begin{align}
\label{11}
P_{\rm con}(r_g^*,\lambda_S)\leq P_{\rm con}\left(\sqrt{\frac{\ln\left(\max\{\mathcal{A}_1,1\}\right)}{\pi\lambda_S}},\lambda_S\right),
\end{align}
where $\mathcal{A}_1$ is defined in Theorem~\ref{thm:Pcon2}.
\end{cor}
\begin{IEEEproof}
The proof follows by observing that $\hat{r}_g$ is the value of $r_g$ that maximizes $P_{\rm con}$ without any constraints, while $r_g^*$ is the value of $r_g$ that maximizes $P_{\rm con}$ with the constraint of $r_g\in\mathcal{G}(\lambda_S)$, as explained in Definition~\ref{def:3}.
\end{IEEEproof}

\subsubsection{Secure Communication Probability}
The secure communication probability, formally defined in Definition~\ref{def:2}, is derived in the following theorem.
\begin{theorem}[Secure communication probability]
\label{thm:Psec}
For a given value of $r_g$ and $\lambda_S$, the probability of secure communication is
\begin{align}
\label{12}
P_{\rm sec}(r_g,\lambda_S)=\exp\left(-2\pi\lambda_S\int_{r_g}^\infty \exp\left(-\frac{\sigma_S^2\beta_Sr_y^{\alpha}}{P_t}\right)\mathcal{L}_{I_2}(\beta_Sr_y^{\alpha})r_y{\rm d}r_y\right),
\end{align}
where $\mathcal{L}_{I_2}(s)=\exp\left(-2\pi\lambda_P P_{\rm active}\int_0^{sr_g^{-\alpha}}\frac{s^{\frac{2}{\alpha}}}{\alpha(1+z)z^{\frac{2}{\alpha}}}{\rm d}z\right)$.
\end{theorem}
\begin{IEEEproof}
See Appendix~\ref{app:Psec}.
\end{IEEEproof}
\begin{remark}
\label{rem:Psec}
The intuitive observations provided in Remark~\ref{teaser} can now be verified using Theorem~\ref{thm:Psec}. 
For instance, the effect of $r_g$ on the distance between the PT and its nearest $ER$ is captured in the integration limits in the above equation. Obviously, as we increase $r_g$, the integration interval will decrease, which increases the value of $P_{\rm sec}$. On the other hand, the effect of $r_g$ on the density of active interferers at the ER is captured in the term $\mathcal{L}_{I_2}(\beta_Sr_x^{\alpha})$. As we mentioned earlier, increasing the value of $r_g$ decreases the interference levels at the ER, which increases the value of ${\rm SINR_S}$, which eventually reduces $P_{\rm sec}$. This is verified here by observing that $\mathcal{L}_{I_2}(s)$ is an increasing function of $r_g$.
\end{remark}
Due to the complexity of the expression provided in Theorem~\ref{thm:Psec} for $P_{\rm sec}$, it is not straightforward to derive expressions for either $\mathcal{G}(\lambda_S)$ or $r_g^*$. However, using the results for $P_{\rm con}$ and $P_{\rm sec}$ in Theorems~\ref{thm:Pcon1} and~\ref{thm:Psec}, it is easy to compute $r_g^*$ numerically. In order to better understand the behavior of the system, we will provide expressions for $P_{\rm con}$ and $P_{\rm sec}$ in both noise-limited as well as interference-limited regimes. In the discussion section, results for both regimes will be compared with the results of the system at different values of $r_g$ resulting in several meaningful insights. In the case of a noise limited regime, the interference terms in the expressions of both $P_{\rm con}$ and $P_{\rm sec}$ will be removed leading to the following corollary.
\begin{cor}
\label{cor:noiselim}
In the noise limited regime, $P_{\rm con}$ will be a decreasing function of $r_g$, while $P_{\rm sec}$ will be an increasing function of $r_g$ as follows
\begin{align}
\label{13}
P_{\rm con}^{\rm Noise\ limited}&=\exp\left(-\left[\lambda_S\pi r_g^2+\beta_P\frac{\sigma_P^2}{P_t}r_1^{\alpha}\right]\right),\\
\label{14}
P_{\rm sec}^{\rm Noise\ limited}&=\exp\left(-2\pi\lambda_S\int_{r_g}^\infty \exp\left(-\frac{\sigma_S^2\beta_Sr_y^{\alpha}}{P_t}\right)r_y{\rm d}r_y\right)\nonumber\\
&=\exp\left(-\frac{2\pi\lambda_S}{\alpha}\left(\frac{P_{t}}{\sigma_S^2\beta_S}\right)^{\frac{2}{\alpha}}\Gamma\left(\frac{2}{\alpha},\frac{r_g^{\alpha}\beta_S\sigma_S^2}{P_{\rm t}}\right)\right).
\end{align}
And the value of $r_g^*$ is presented by the following equation
\begin{align}
\label{15}
\Gamma\left(\frac{2}{\alpha},\frac{(r_g^*)^{\alpha}\beta_S\sigma_S^2}{P_{\rm t}}\right)&=\min\left\{\frac{\alpha{\rm Ln}\left(\frac{1}{\epsilon}\right)}{2\pi\lambda_{S}\left(\frac{P_{t}}{\sigma_S^2\beta_S}\right)^{\frac{2}{\alpha}}},\Gamma\left(\frac{2}{\alpha}\right)\right\}.
\end{align}
\end{cor}
\begin{IEEEproof}
Eq.~\ref{13} and~\ref{14} follow directly by removing the interference terms from the results in Theorem~\ref{thm:Pcon1} and Theorem~\ref{thm:Psec}. Since $P_{\rm con}^{\rm Noise\ limited}$ is a decreasing function of $r_g$, and $P_{\rm sec}^{\rm Noise\ limited}$ is an increasing function of $r_g$, the primary network picks the minimum value of $r_g$ that satisfies the condition $P_{\rm sec}\geq\epsilon$. Substituting for $P_{\rm sec}$ using Eq.~\ref{14} in the inequality $P_{\rm sec}\geq\epsilon$ leads to Eq.~\ref{15}.
\end{IEEEproof}
In the case of an interference-limited regime, the noise terms in the expressions of both $P_{\rm con}$ and $P_{\rm sec}$ will be ignored leading to the following corollary.
\begin{cor}
\label{cor:interferencelim}
In the interference limited regime, expressions for $P_{\rm con}$ and $P_{\rm sec}$ are provided below.
\begin{align}
\label{101}
P_{\rm con}^{\rm Int.\ limited}&=\exp\left(-\left[\lambda_S\pi r_g^2+\frac{2\pi^2\lambda_P P_{\rm active}\beta_P^{\frac{2}{\alpha}}r_1^{2}}{\alpha\sin(\frac{2}{\alpha}\pi)}\right]\right),\\
\label{102}
P_{\rm sec}^{\rm Int.\ limited}&=\exp\left(-2\pi\lambda_S\int_{r_g}^\infty \mathcal{L}_{I_2}(\beta_Sr_x^{\alpha})r_x{\rm d}r_x\right).
\end{align}
\end{cor}
\begin{IEEEproof}
These results follow by substituting for $\sigma_P^2=\sigma_S^2=0$ in Theorems~\ref{thm:Pcon1} and~\ref{thm:Psec}. 
\end{IEEEproof}
The main take-away from this subsection is that each of $P_{\rm con}$ and $P_{\rm sec}$ are functions of $\lambda_S$, which consequently implies that $r_g^*$ is a function of $\lambda_S$. In the next subsection, we will show that optimal density $\lambda_S^*$ selected by the secondary network is also a function of $r_g$. 
\subsection{Secondary Network}
The objective of the secondary network is to maximize the density of successfully powered ERs $P_{\rm energy}$, introduced in Eq.~\ref{9}, which is derived in the following Theorem.
\begin{theorem}[Density of successfully powered ERs]
\label{thm:Penergy}
For a given value of $r_g$ and $\lambda_S$, the density of successfully powered ERs is
\begin{align}
\label{16}
P_{\rm energy}(r_g,\lambda_S)=\lambda_S\int_{r_g}^{\infty}2\pi\lambda_P P_{\rm active}r_p\exp\left(-\pi\lambda_P P_{\rm active}(r_p^2-r_g^2)-\frac{E_{\rm min}r_p^{\alpha}}{P_t\eta} \right){\rm d}r_p.
\end{align}
\end{theorem}
\begin{IEEEproof}
See Appendix~\ref{app:Penergy}.
\end{IEEEproof}
The above equation is a product of two terms: (i) the density of ERs $\lambda_S$, and (ii) the integral term, which is the expression derived for the probability $\mathbb{P}(E_{H}\geq E_{\rm min})$. Consequently, we can claim the existence of an optimal value of $\lambda_S$ that maximizes the density of successfully powered ERs $P_{\rm energy}$. The reason behind that is the dependence of the density of active sources of RF energy (active PTs) on the density of ERs. This interesting trade-off arises due to the use of secrecy guard zones by the primary network. Obviously, the value of $P_{\rm energy}$ is a function of $r_g$. Hence, the value of $\lambda_S^*$, that maximizes $P_{\rm energy}$, is also a function of $r_g$. Unfortunately, $\lambda_S^*$ can not be computed from the above expression due to its complexity. Hence, to get more insights on the relation between $\lambda_S^*$ and $r_g$, we derive a lower bound on $P_{\rm energy}$ in the following corollary. We use this lower bound to compute $\lambda_S^*$ as a function of $r_g$.
\begin{cor}
The value of $P_{\rm energy}$ is lower bounded as follows
\begin{align}
\label{111}
P_{\rm energy}\geq \lambda_S\int_{r_g}^{\infty}2\pi\lambda_P P_{\rm active}r_p\exp\left(-\pi\lambda_P(r_p^2-r_g^2)-\frac{E_{\rm min}r_p^{\alpha}}{P_t\eta} \right){\rm d}r_p.
\end{align}
The value of $\lambda_S$ that maximizes the lower bound is
\begin{align}
\label{112}
\lambda_S^*=\frac{1}{\pi r_g^2}.
\end{align}
\end{cor}
\begin{IEEEproof}
The lower bound follows by simply replacing $P_{\rm active}$ inside the exponent in Eq.~\ref{16} with unity. The value of $\lambda_S^*$ follows by differentiating Eq.~\ref{111} with respect to $\lambda_S$. 
\end{IEEEproof}
\begin{remark}
\label{rem:energy}
It can be noted from Eq.~\ref{112} that the value of $\lambda_S^*$ is a decreasing function of $r_g$. This agrees with intuition since when the value of $r_g$ increases, the secondary network needs to decrease $\lambda_S$ in order to maintain the density of active PTs, which are the sources of RF energy.
\end{remark}
 The mutual coupling between $r_g^*$ and $\lambda_S$ as well as $\lambda_S^*$ and $r_g$ can be best modeled using tools from game theory. This will be the core of the next section.
\section{Game Theoretical Modeling}
\label{sec:game}
Building on all the insights and comments given in the previous section, we can model the interaction between the two networks using tools from game theory. For a given value of $\lambda_S$, the primary network selects $r_g^*$ as presented in Definition~\ref{def:3}, which can be rewritten as
\begin{align}
\label{17}
r_g^*=\arg\underset{r_g\geq0}{\max}\ P_{\rm con}(r_g,\lambda_S)\mathbbm{1}(P_{\rm sec}(r_g,\lambda_S)\geq\epsilon),
\end{align}
where $\mathbbm{1}(\Xi)=1$ if the condition $\Xi$ is satisfied, and equals zero otherwise. On the other hand, for a given value of $r_g$, the secondary network selects $\lambda_S^*$ as follows
\begin{align}
\label{18}
\lambda_S^*=\arg\underset{\lambda_S\geq0}{\max}\ P_{\rm energy}(r_g,\lambda_S).
\end{align}
Observing Eq.~\ref{17} and~\ref{18}, it is fairly straightforward to see that the system setup can be modeled as a non-cooperative static game with two players: (i) the primary network is {\em player 1}, and (ii) the secondary network is {\em player 2}. The utility function of player 1 is
\begin{align}
\label{19}
U_1(r_g,\lambda_S)=P_{\rm con}(r_g,\lambda_S)\mathbbm{1}(P_{\rm sec}(r_g,\lambda_S)\geq\epsilon),
\end{align}
while the utility function of player 2 is 
\begin{align}
\label{20}
U_2(r_g,\lambda_S)=P_{\rm energy}(r_g,\lambda_S).
\end{align}
For this game, the main objective is to find the values of $r_g^*$ and $\lambda_S^*$ where each of the two networks has no tendency to change its tuning parameter. This is the definition of Nash equilibrium (NE). This can be mathematically modeled as
\begin{align}
\label{21}
r_g^*&=\arg\underset{r_g\geq0}{\max}\ P_{\rm con}(r_g,\lambda_S^*)\mathbbm{1}(P_{\rm sec}(r_g,\lambda_S^*)\geq\epsilon),\nonumber\\
\lambda_S^*&=\arg\underset{\lambda_S\geq0}{\max}\ P_{\rm energy}(r_g^*,\lambda_S).
\end{align}
Unfortunately, due to the complexity of the expressions of $P_{\rm con}$ in Theorem~\ref{thm:Pcon1} and $P_{\rm sec}$ in Theorem~\ref{thm:Psec}, it is challenging to provide a closed-form solution for NE. However, based on the comments given in Remark~\ref{teaser_2} and the result in Theorem~\ref{thm:Pcon2}, we provide a learning algorithm that assists both networks to find NE.

The proposed learning algorithm is a simple best-response based algorithm~\cite{sandholm2010population,6868232,barron2010best}. In each iteration, each network updates its parameters according to the opponent's network parameter in the previous iteration. This can be mathematically presented as follows:
\begin{align}
\label{22}
r_g^{(n)}&=r_g^*\left(\lambda_S^{(n-1)}\right),\nonumber\\
\lambda_S^{(n)}&=\lambda_S^*\left(r_g^{(n-1)}\right),
\end{align} 
where $r_g^{(n)}$ and $\lambda_S^{(n)}$ are the outputs of the algorithm in the $n$-th iteration, $r_g^*\left(\lambda_S^{(n-1)}\right)$ is computed using Eq.~\ref{17} for $\lambda_S=\lambda_S^{(n-1)}$, and $\lambda_S^*\left(r_g^{(n-1)}\right)$ is computed using Eq.~\ref{18} for $r_g=r_g^{(n-1)}$. Assuming that the secondary network has a maximum possible deployment density, $\lambda_S\leq\lambda_{S,{\rm max}}$, the algorithm is provided next.\\\vspace{2mm}\\
    \begin{tabular}{  p{13cm} }
    \hline\hline
    \textbf{Algorithm}: Proposed algorithm for finding the NE network parameters $r_g^*$ and $\lambda_S^*$  \\ 
    \textbf{input}: $\lambda_P$, $\lambda_{S,{\rm max}}$, $\beta_P$, $\beta_S$, $P_t$, $\alpha$, $\eta$, and $E_{\rm min}$.\\
    \textbf{output}: $r_g^{*}$, $\lambda_S^{*}$.\\ \hline
    \textbf{Initialization}: $r_g^{(0)}=0$ for all PTs in $\Phi_{\rm P}$, $\lambda_S^{(0)}=\lambda_{S,{\rm max}}$, $n=1$\\ \hline
1: \textbf{Repeat}\\
2: \ \ \ \ \ $r_g^{(n)}=r_g^*\left(\lambda_S^{(n-1)}\right)$\\
3: \ \ \ \ \ $\lambda_S^{(n)}=\lambda_S^*\left(r_g^{(n-1)}\right)$\\
4: \ \ \ \ \ $n=n+1$\\
5: \textbf{Until} $r_g^{(n)}=r_g^{(n-1)}$ $\&$ $\lambda_S^{(n)}=\lambda_S^{(n-1)}$\\ \hline\hline
    \end{tabular}
    \vspace{6mm}\\
    
Note that we assume the ability of each network to estimate perfectly the opponent's action in the previous iteration. Including the possibility of the estimation error and its effect on convergence is left as a promising direction of future work. In the next section, we will show using simulations the convergence of the proposed algorithm to the NE of the modeled game. 
\section{Results and Discussion}
\begin{figure}
\centering
\includegraphics[width=0.5\columnwidth]{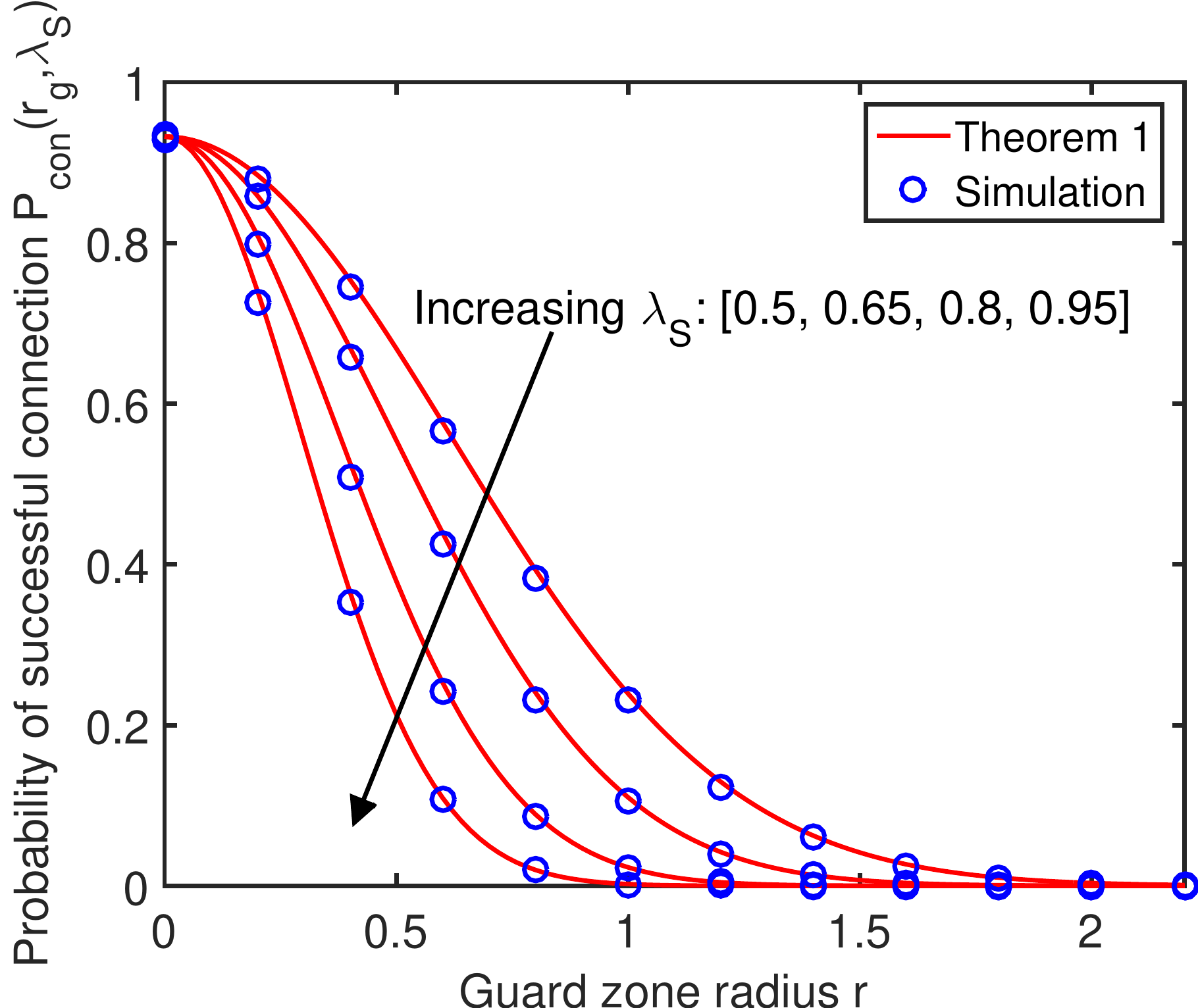}
\caption{The effect of $\lambda_S$ on the behavior of $P_{\rm con}$ against different values of $r_g$. The SNR value at the legitimate receiver is $\gamma_P=7$ dB.}
\label{fig:1}
\end{figure}
In this section, we will use both theoretical and simulation results (obtained from Monte-Carlo trials) to analyze the performance of both primary and secondary networks. The values of the system parameters used for the simulation setup are: $\lambda_P=1$, $\eta=0.75$, $E_{\rm min}=10^{-4}$ Joules, $\alpha=4$, $\lambda_{S,{\rm max}}=2$, $P_t=1$, $\beta_P=3$ dB, $\beta_S=0$ dB, $r_1=0.1$, and $\epsilon=0.8$. The values of the rest of the parameters will be defined either on the figures or in their captions. We also refer to the SNR values at the primary and secondary receivers as $\gamma_P=\frac{P_t}{\sigma_S^2}$ and $\gamma_S=\frac{P_t}{\sigma_S^2}$ respectively. In addition, we define the ratio of $\lambda_S$ to its maximum value by $\delta_S=\frac{\lambda_S}{\lambda_{S,{\rm max}}}$. Our main goals in this section are: (i) to validate the approximations used in our derivations, (ii) to get further insights on the behavior of both the networks, and (iii) to verify the comments provided in the Remarks throughout the paper. We first study the successful connection probability $P_{\rm con}$ in Fig.~\ref{fig:1}. First note that the simulation parameters chosen above are such that we get $\mathcal{A}_1 < 1$, where  $\mathcal{A}_1$ is defined in Theorem~\ref{thm:Pcon2}. The results in Fig.~\ref{fig:1} show that increasing the density of ERs decreases $P_{\rm con}$. Although increasing $\lambda_S$ decreases the density of active interferers (which should increase $P_{\rm con}$), it also decreases the probability of the PT being active (which decreases $P_{\rm con}$). Ultimately, for this setup, $P_{\rm con}$ is a decreasing function of $\lambda_S$ because of having $\mathcal{A}_1< 1$. As explained in Remark~\ref{rem:Pcon2}, $\mathcal{A}_1< 1$ leads to a noise limited system from the perspective of the successful connection probability, which means that the effect of $\lambda_S$ on the density of active interferers is negligible compared to its effect on the probability of the PT being active. Further, note that because of having $\mathcal{A}_1<1$, $P_{\rm con}$ is also a decreasing function of $r_g$. As a result, $r_g^*$ will be the minimum value of $r_g$ that ensures $P_{\rm sec}\geq\epsilon$. The value of $P_{\rm sec}$ as a function of $r_g$ is plotted in Fig.~\ref{fig:2} for different values of $\gamma_S$. Consistent with the intuition, increasing $\gamma_S$ decreases $P_{\rm sec}$. We also note that at low values of $r_g$ the effect of $\gamma_S$ is hardly noticeable, while at higher values of $r_g$, the effect of $\gamma_S$ becomes significant. Furthermore, we note that $r_g^*$ is an increasing function of $\gamma_S$. To better understand the effect of $\gamma_S$ on the behavior of $P_{\rm sec}$, we plot in Fig.~\ref{fig:4} the value of $P_{\rm sec}$ for both the noise limited and the interference limited regimes, which are derived in Corollary~\ref{cor:noiselim} and~\ref{cor:interferencelim}, respectively. These results lead to the following key insights: 
\begin{figure}[!t]
    \centering
    \begin{minipage}{.45\textwidth}
        \centering
\includegraphics[width=1\columnwidth]{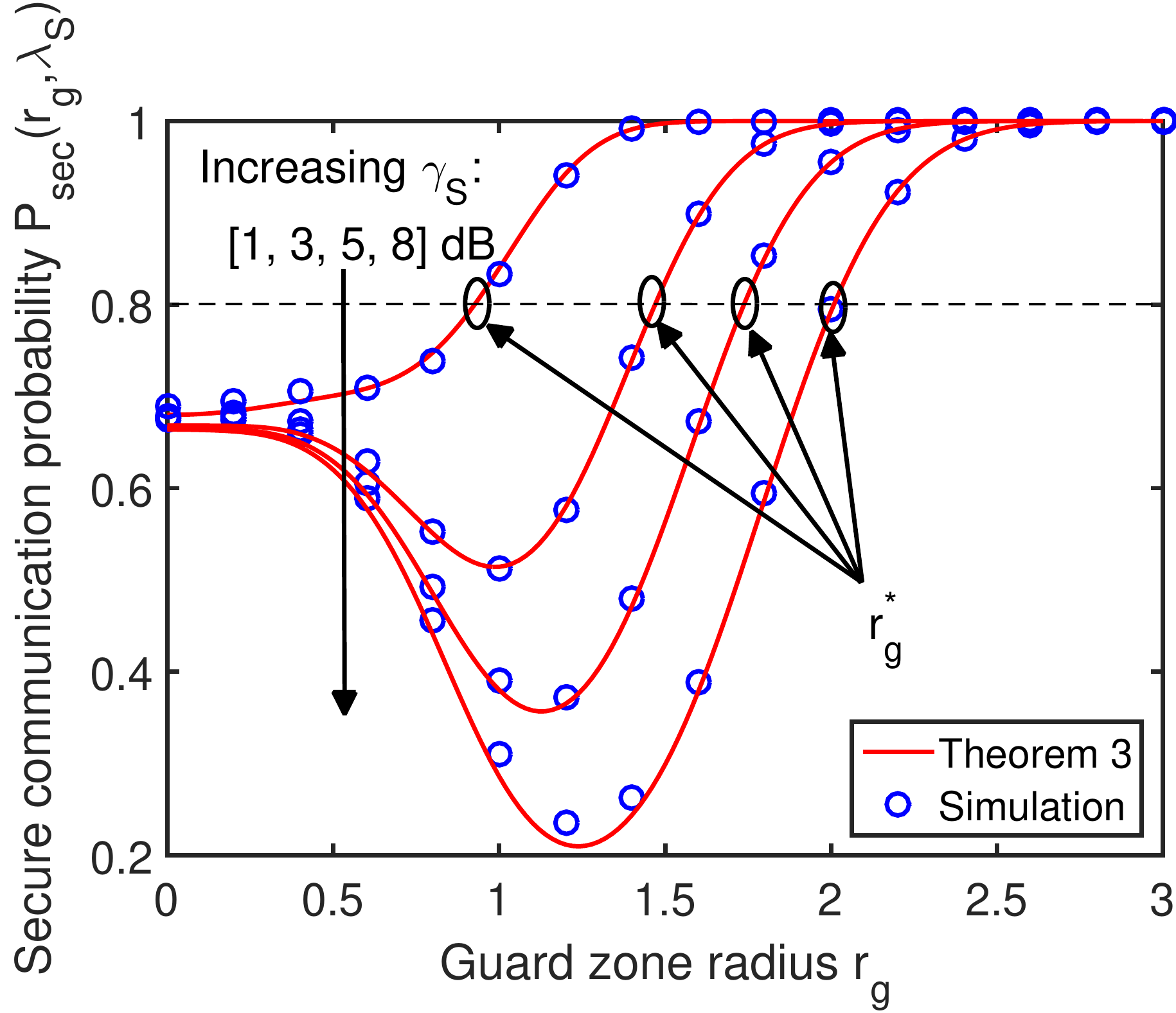}
\caption{The effect of $\gamma_S$ on the behavior of $P_{\rm sec}$ against different values of $r_g$. The density of ERs is $\lambda_S=0.6$.}
\label{fig:2}
    \end{minipage}%
    \hfill
    \begin{minipage}{0.45\textwidth}
        \centering
\includegraphics[width=1\columnwidth]{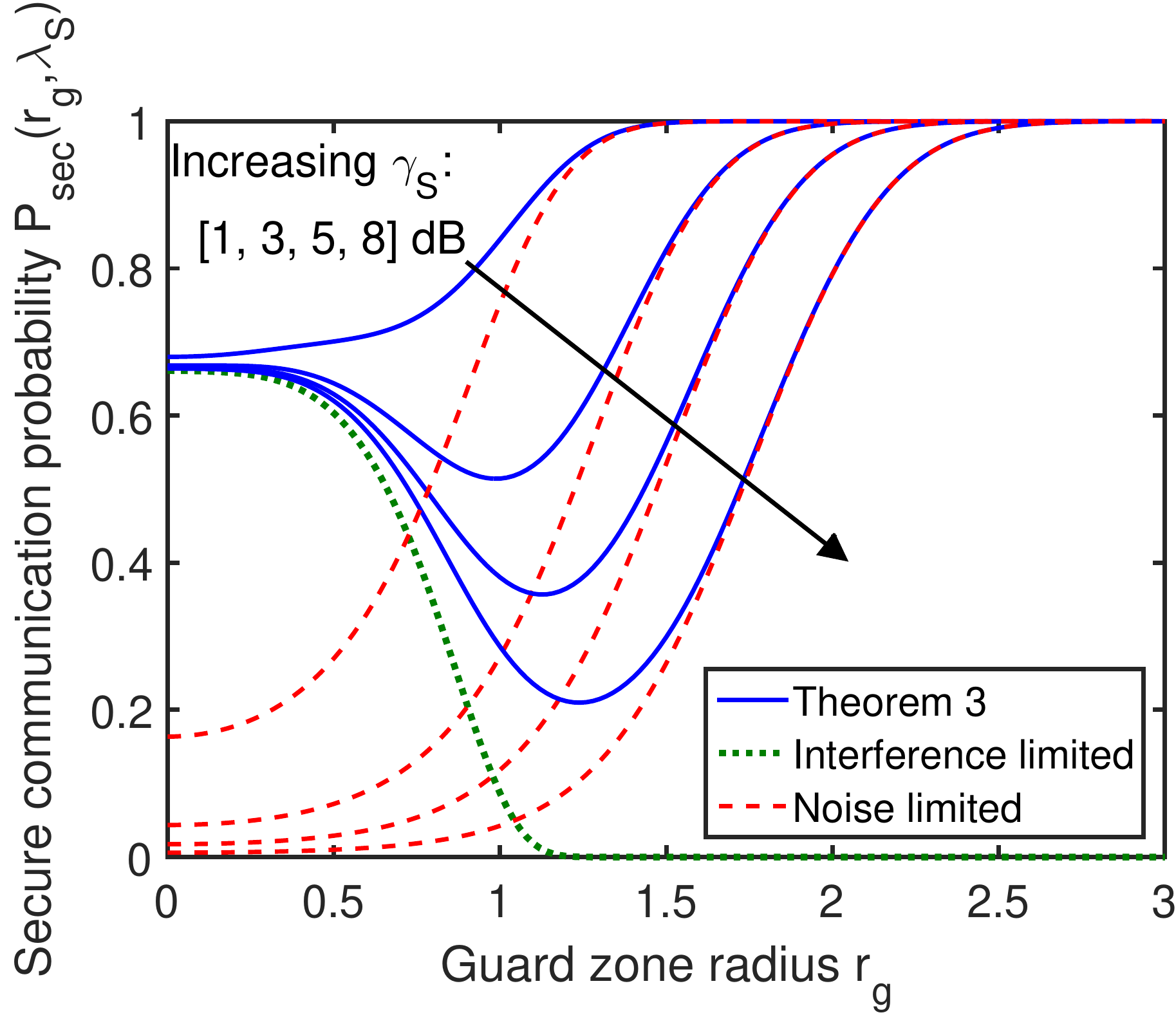}
\caption{Comparing $P_{\rm sec}$, $P_{\rm sec}^{\rm Noise Limited}$, and $P_{\rm sec}^{\rm Int. Limited}$ as functions of $r_g$ for different values of $\gamma_S$. The density of ERs is $\lambda_S=0.6$.}
\label{fig:4}
    \end{minipage}
\end{figure}

\begin{itemize}
\item At lower values of $r_g$, the density of active interferers is relatively high. Hence, we observe that from the secure communication probability perspective, the system is interference limited in this regime. Furthermore, increasing the SNR value $\gamma_S$ has a negligible effect. This is a direct consequence of being in the interference-limited regime. 
\item At higher values of $r_g$, the density of active interferers decreases, which drives the system to the noise limited regime. As a result, the noise power has more noticeable effect on $P_{\rm sec}$. Equivalently, increasing value of $\gamma_S$ decreases the value of $P_{\rm sec}$ at higher values of $r_g$. 
\item Obviously, $P_{\rm sec}^{\rm Int.\ Limited}$ is a decreasing function of $r_g$, while $P_{\rm sec}^{\rm Noise\ Limited}$ is an increasing function of $r_g$. However, the behavior of $P_{\rm sec}$ is harder to describe. We note the existence of a local minimum of $P_{\rm sec}$ below which it behaves more like $P_{\rm sec}^{\rm Int.\ Limited}$, whereas above this local minimum, $P_{\rm sec}$ behaves similar to $P_{\rm sec}^{\rm Noise\ Limited}$.
\end{itemize}

\begin{figure}[!t]
    \centering
    \begin{minipage}{.45\textwidth}
        \centering
\includegraphics[width=1\columnwidth]{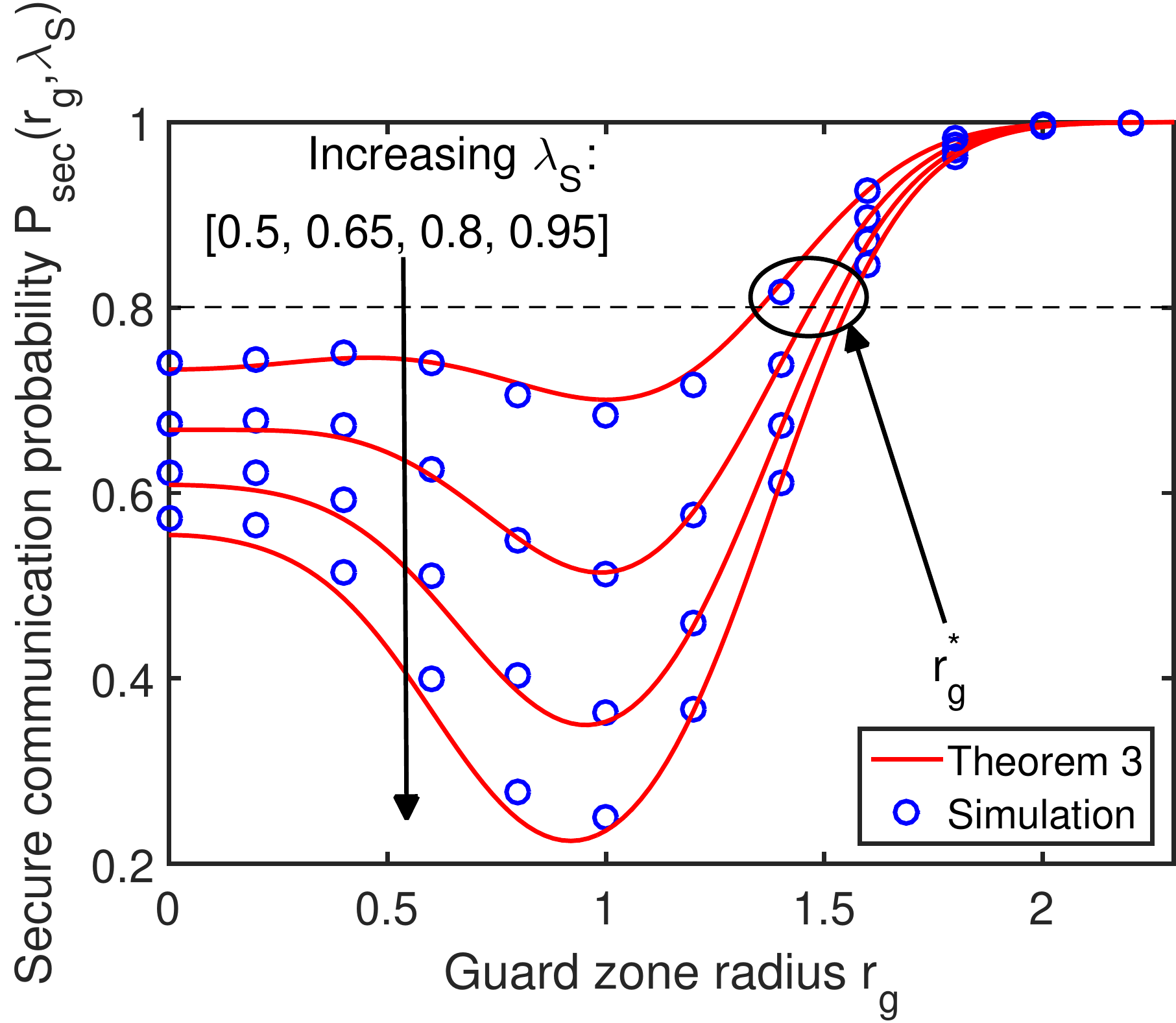}
\caption{The effect of $\lambda_S$ on the behavior of $P_{\rm sec}$ against different values of $r_g$. The SNR value at ERs is $\gamma_S=4.8$ dB.}
\label{fig:3}
    \end{minipage}%
    \hfill
    \begin{minipage}{0.45\textwidth}
        \centering
\includegraphics[width=1\columnwidth]{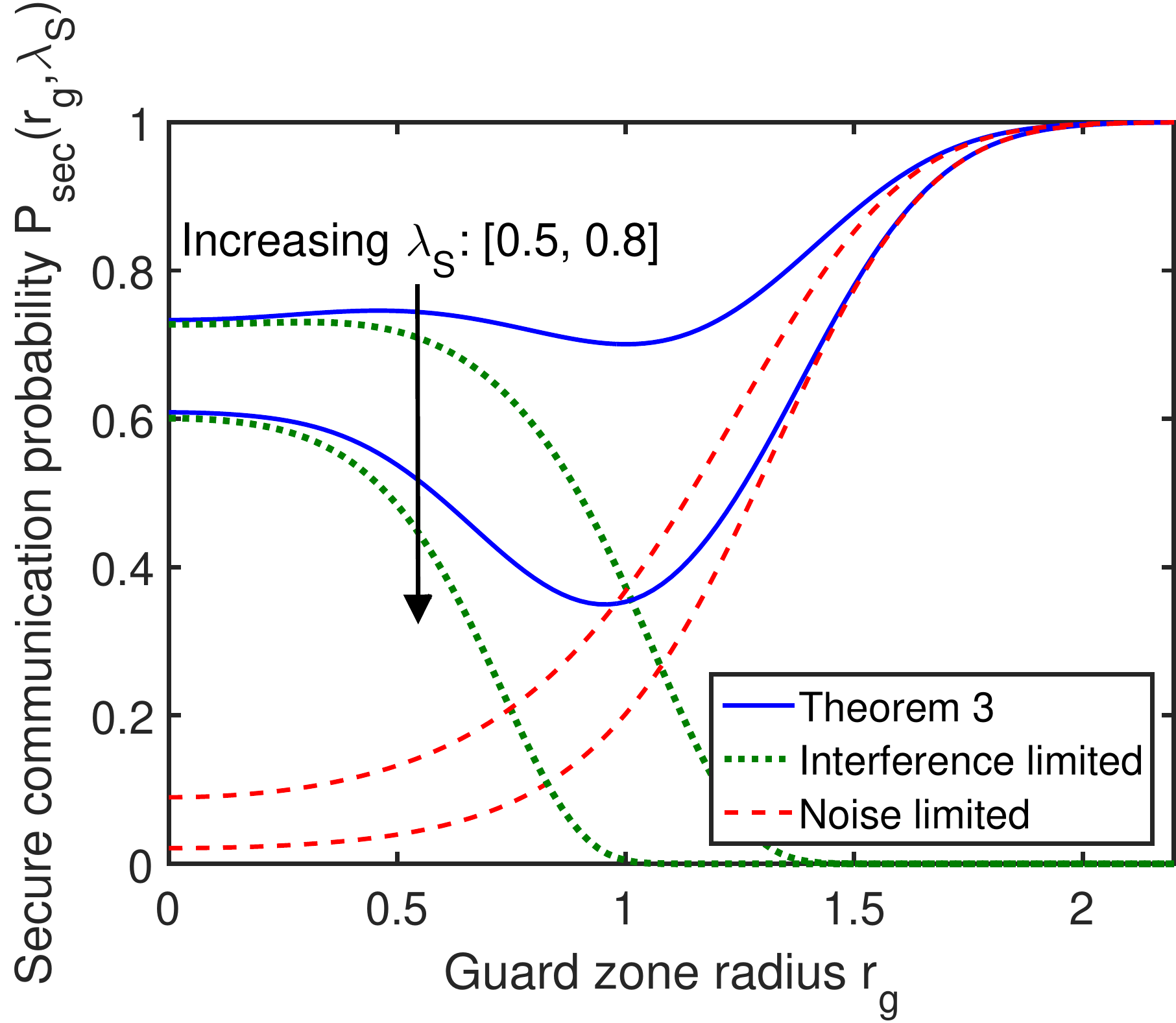}
\caption{Comparing $P_{\rm sec}$, $P_{\rm sec}^{\rm Noise Limited}$, and $P_{\rm sec}^{\rm Int. Limited}$ as functions of $r_g$ for different values of $\lambda_S$. The SNR value at ERs is $\gamma_S=4.8$ dB.}
\label{fig:5}
    \end{minipage}
\end{figure}

In Fig.~\ref{fig:3}, we study the effect of $\lambda_S$ on the behavior of $P_{\rm sec}$ against different values of $r_g$. We note that $r_g^*$ increases with $\lambda_S$. We also note that, unlike $\gamma_S$, the effect of $\lambda_S$ is more prominent at the lower values of $r_g$. As explained above, the reason is that the system is interference limited at lower values of $r_g$, where increasing $\lambda_S$ decreases the density of interferers, which in turn decreases $P_{\rm sec}^{\rm Int.\ Limited}$. We plot $P_{\rm sec}$, $P_{\rm sec}^{\rm Int.\ Limited}$, and $P_{\rm sec}^{\rm Noise\ Limited}$ in Fig.~\ref{fig:5} to verify these arguments. We also note that at higher values of $r_g$, as we stated earlier, the system is noise limited, where $\lambda_S$ has less effect on $P_{\rm sec}^{\rm Noise\ Limited}$ compared to $P_{\rm sec}^{\rm Int.\ Limited}$. This can be verified from Fig.~\ref{fig:3} by observing that the gaps between the $P_{\rm sec}$ curves for different values of $\lambda_S$ get tighter as $r_g$ increases.

\begin{figure}
\centering
\includegraphics[width=0.5\columnwidth]{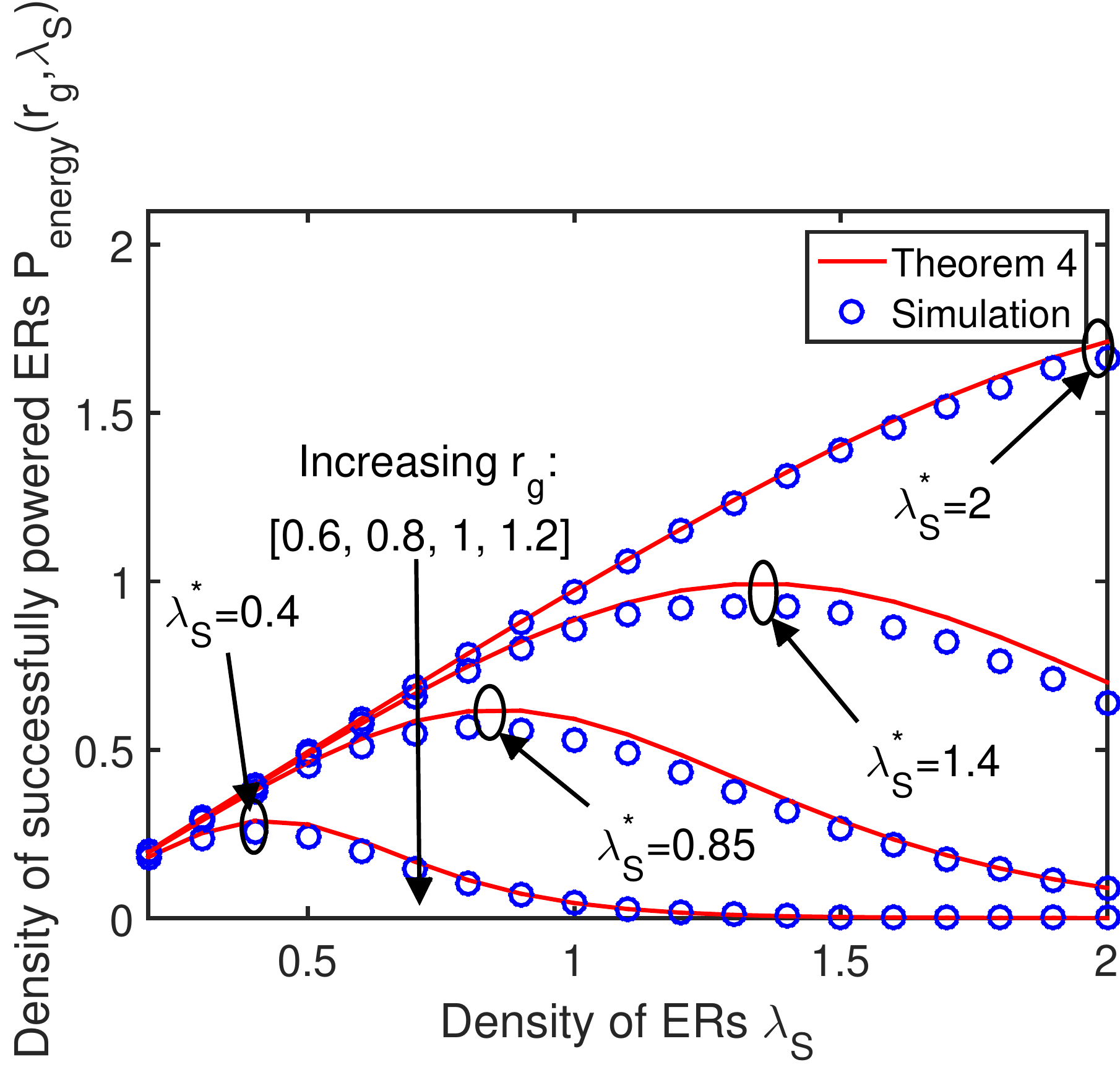}
\caption{The density of successfully powered ERs $P_{\rm energy}$ against $\lambda_S$ for different values of $r_g$. }
\label{fig:6}
\end{figure}

Now that we have better understanding of the behavior of $P_{\rm con}$ and $P_{\rm sec}$, we study the energy harvesting performance of the secondary network. In Fig.~\ref{fig:6}, we illustrate the behavior of $P_{\rm energy}$ against $\lambda_S$ for different values of $r_g$. We note that at lower values of $r_g$, the optimal value of $\lambda_S$ is its maximum value $\lambda_{S,{\rm max}}$. This is consistent with intuition that at lower values of $r_g$, the density of active PTs (sources of RF energy) will hardly get affected by increasing $\lambda_S$. As $r_g$ increases, the impact of $\lambda_S$ on the density of active PTs becomes noticeable. As shown in Fig.~\ref{fig:6}, this eventually leads to the existence of a local maximum of $P_{\rm energy}$. We also note that as the value of $r_g$ increases, the optimal value $\lambda_S^*$ decreases. This is also consistent with our comments in Remark~\ref{rem:energy} that as $r_g$ increases, the secondary network needs to decrease its density in order to maintain the density of active PTs that provide the RF energy to the ERs. 

\begin{figure}[!t]
    \centering
    \begin{minipage}{.45\textwidth}
        \centering
	\includegraphics[width=1\columnwidth]{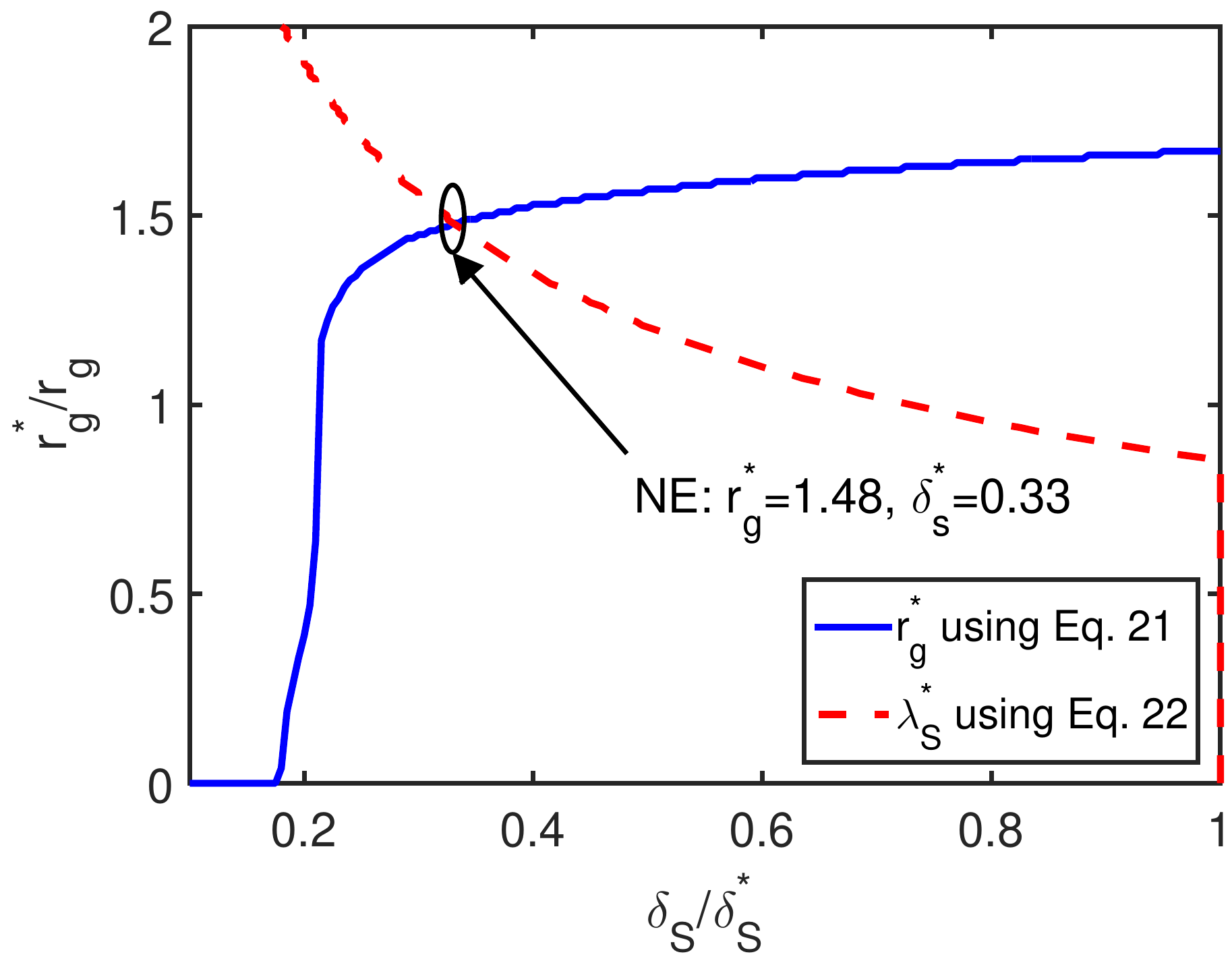}
\caption{Plotting both $r_g^*$ against $\delta_S$ and $\delta_S^*$ against $r_g$ on the same plot to compute the NE values.}
\label{fig:7}
    \end{minipage}%
    \hfill
    \begin{minipage}{0.45\textwidth}
        \centering
\includegraphics[width=1\columnwidth]{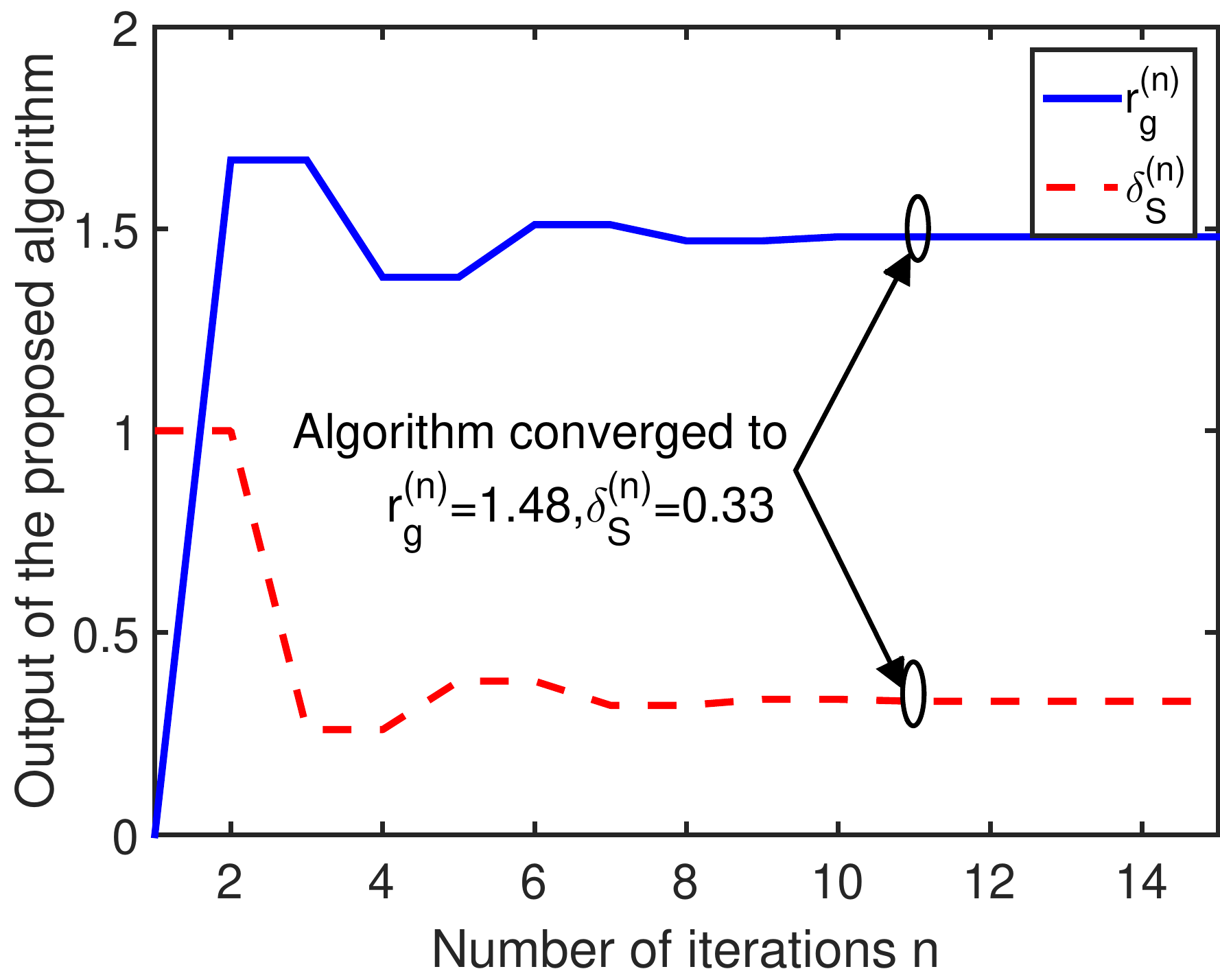}
\caption{The proposed algorithm converges to Nash equilibrium in a finite number of iterations.}
\label{fig:8}
    \end{minipage}
\end{figure}

Taking a second look at Figs.~\ref{fig:3} and~\ref{fig:6}, we can easily verify our earlier comments that the parameters selected by each of the networks ($r_g$ for the primary network and $\lambda_S$ for the secondary network) affect the value of the optimal parameter of the other network. Using the procedure provided in Sec.~\ref{sec:game}, we simulate the interaction between both networks by simulating the relations between $r_g^*$ and $\delta_S$ as well as $\delta_S^*$ and $r_g$, where $\delta_S=\frac{\lambda_S}{\lambda_{S,{\rm max}}}$ is the normalized value of the density of ERs. In Fig.~\ref{fig:7}, we plot $r_g^*$ (on the y-axis) for different values of $\delta_S$ (on the x-axis). On the same figure, we plot $\delta_S^*$ (on the x-axis) for different values of $r_g$ (on the y-axis). The intersection of both curves represents the value of NE where each of the two networks has no intention to deviate as explained in Eq.~\ref{21}. In Fig.~\ref{fig:8}, we evaluate the performance of the algorithm proposed in Sec.~\ref{sec:game}. The results demonstrate the convergence of the proposed algorithm to the NE found from Fig.~\ref{fig:7} after less than 13 iterations.

\section{Conclusions}

The emergence of IoT regime is characterized by the deployment of billions of devices some of which may be equipped with energy harvesting capability. Due to the ubiquity of RF signals, harvesting energy from the ambient RF signals is perhaps the most attractive option for such devices. This raises the possibility of some of these devices acting as eavesdroppers, which motivates the need to study secure wireless power transfer to energy receivers acting as potential eavesdroppers. While this problem received some attention in the literature, the existing works are limited to simple point-to-point or simple deterministic topologies, which are not sufficient to accurately model the massive scale of IoT. In this paper, we developed the first comprehensive stochastic geometry-based model to study the performance of an ambient RF energy harvesting network (secondary network) when the sources of RF signals are transmitters with secrecy guard zones (primary network). First, using tools from stochastic geometry, we derived the successful connection and secure communication probabilities of the primary network. Next, we derived the density of successfully powered nodes in the secondary network. Furthermore, we showed that the performance metrics of the two networks are coupled. In particular, we showed that the optimal guard zone radius (that maximizes the successful connection probability while maintaining the secure communication probability above a predefined threshold) is a function of the deployment density of the secondary network. In addition, we showed that the optimal deployment density of the secondary network (that maximizes the density of successfully powered nodes) is a function of the guard zone radius of the primary network. Hence, we used tools from game theory to model this interesting coupling between the two networks. In particular, we showed that such system can be modeled as a two player non-cooperative game. In addition, we proposed a best-response based algorithm and demonstrated with simulations its convergence to Nash equilibrium.

This paper is one of the few concrete works that symbiotically merge tools from stochastic geometry and game theory. It can be extended in many directions. From the secrecy perspective, it will be useful to extend the proposed model to incorporate other secrecy enhancing techniques, such as beamforming and artificial noise technique. From the modeling perspective, it is worthwhile to investigate other meaningful morphologies, such as the one in which ERs are clustered around the RF sources. From game theory perspective, it will be interesting to consider the effect of imperfect estimation of the opponent's action on the performance of the proposed algorithm. 

\appendices
\section{Proof of Theorem~\ref{thm:Pcon1}}\label{app:Pcon}
Recalling the expression for ${\rm SINR_P}$ given in Eq.~\ref{2}, specifically the indicator function $\delta_i$ that indicates which interferer is active and which is silent, we concluded that the locations of active PTs can be modeled by PHP $\Psi$ in Eq.~\ref{3}. However, before using $\Psi$ in our analysis, we need to make it clear that $\delta_i$ for different $x_i\in\Phi_P$ are correlated. This implicit correlation arises from the dependence of $\delta_i$ for all $i$ on the PPP $\Phi_S$. However, capturing this correlation in our analysis will significantly reduce the tractability of the results. Hence, this correlation will be ignored in our analysis for the sake of tractability. The tightness of this approximation will be verified in the Numerical Results section. Now revisiting the expression of $P_{\rm con}$ in Eq.~\ref{4}, we note that the correlation between $\delta_1$ at the typical PT and $\delta_i$ values at each of the interferers in the expression of ${\rm SINR_P}$ is the only source of correlation between the events $(R_e\geq r_g)$ and $({\rm SINR_P}\geq\beta_P)$. Hence, ignoring this correlation will lead to the following
\begin{align}
\label{app:1}
P_{\rm con}=\mathbb{P}(R_e\geq r_g)\mathbb{P}({\rm SINR_P}\geq\beta_P).
\end{align}
The first term in the above expression represents $P_{\rm active}=\exp\left(-\pi\lambda_Sr_g^2\right)$ (please recall Eq.~\ref{1} where ${P}_{\rm active}$ was derived). To derive the second term in the above expression, characterizing the statistics of the interference from a PHP modeled network at a randomly located reference point (the typical PR) is required. However, ignoring the correlation between $\{\delta_i\}$, for the sake of tractability as explained above, is equivalent to approximating the PHP $\Psi$ with a PPP $\Psi_P$ of equivalent density $\tilde{\lambda}_P=\lambda_PP_{\rm active}$. Defining $I=\sum_{x_i\in\Psi_P}h_i\|x_i\|^{-\alpha}$, then
\begin{align}
\label{app:2}
\mathbb{P}({\rm SINR_P}\geq\beta_P)&=\mathbb{P}\left(\frac{h_1r_1^{-\alpha}}{I+\frac{\sigma_P^2}{P_t}}\geq\beta_P\right)\stackrel{(a)}{=}\mathbb{E}_I\left[\exp\left(-\beta_P\left(I+\frac{\sigma_P^2}{P_t}\right)r_1^{\alpha}\right)\right]\nonumber\\
&{=}\exp\left(-\beta_P\frac{\sigma_P^2}{P_t}r_1^{\alpha}\right)\mathbb{E}_I\left[\exp\left(-\beta_P I r_1^{\alpha}\right)\right]\stackrel{(b)}{=}\exp\left(-\beta_P\frac{\sigma_P^2}{P_t}r_1^{\alpha}\right)\mathcal{L}_I\left(\beta_P r_1^{\alpha}\right),
\end{align}
where $h_1\sim \exp(1)$ leads to step (a), and in step (b) we use the definition of Laplace transform of $I$ which is $\mathcal{L}_I(s)=\mathbb{E}\left[\exp\left(-sI\right)\right]$. The Laplace transform of the interference in PPP is a well established result in the literature~\cite{AndGupJ2016}. For completeness, the derivation of $\mathcal{L}_I(s)$ is provided next.
\begin{align}
\label{app:3}
\mathcal{L}_I(s)&=\mathbb{E}_{\Psi_P,\{h_i\}}\left[\exp\left(-s\sum_{x_i\in\Psi_P}h_i\|x_i\|^{-\alpha}\right)\right]=\mathbb{E}_{\Psi_P,\{h_i\}}\left[\prod_{x_i\in\Psi_P}\exp\left(-sh_i\|x_i\|^{-\alpha}\right)\right]\nonumber\\&\stackrel{(c)}{=}\mathbb{E}_{\Psi_P}\left[\prod_{x_i\in\Psi_P}\frac{1}{1+s\|x_i\|^{-\alpha}}\right]\stackrel{(d)}{=}\exp\left(-\tilde{\lambda}_P\int_{x\in\mathbb{R}^2}1-\frac{1}{1+s\|x\|^{-\alpha}}{\rm d}x\right)\nonumber\\
&\stackrel{(e)}{=}\exp\left(-2\pi\tilde{\lambda}_P\int_{0}^{\infty}\frac{sr_x^{-\alpha}}{1+sr_x^{-\alpha}}r_x{\rm d}r_x\right)\stackrel{(f)}{=}\exp\left(-\frac{2\pi^2\tilde{\lambda}_Ps^{\frac{2}{\alpha}}\csc\left(\frac{2\pi}{\alpha}\right)}{\alpha}\right),
\end{align}
where knowing that the set of fading gains ${h_i}$ are i.i.d with $h_i\sim \exp(1)$ leads to step (c), step (d) results from using the probability generating function (PGFL) of PPP~\cite{haenggi2012stochastic}, step (e) results from converting to polar co-ordinates, and step (f) follows after some mathematical manipulations. Substituting Eq.~\ref{app:3} in Eq.~\ref{app:2} and then in Eq.~\ref{app:1} leads to the final result in Theorem~\ref{thm:Pcon1}.
\section{Proof of Theorem~\ref{thm:Pcon2}}\label{app:Pcon2}
Observing the expression of $P_{\rm con}$ in Theorem~\ref{thm:Pcon1}, we note that it can be rewritten as a function of $P_{\rm active}=\exp(-\lambda_S\pi r_g^2)$ as follows
\begin{align}
\label{app:4}
P_{\rm con}=\exp\left(-\beta_P\frac{\sigma_P^2}{P_t}r_1^{\alpha}\right)P_{\rm active}\exp\left(-P_{\rm active}\mathcal{A}_1\right),
\end{align}
where $\mathcal{A}_1=\frac{2\pi^2\lambda_P \beta_P^{\frac{2}{\alpha}}r_1^{2}}{\alpha\sin(\frac{2}{\alpha}\pi)}$. To get more information about the behavior of $P_{\rm con}$ against $P_{\rm active}$, we compute the first derivative (with respect to $P_{\rm active}$). Given that $P_{\rm active}$ is a decreasing function of $r_g$ (recall Eq.~\ref{1}), we conclude the following
\begin{enumerate}
\item If $1-\mathcal{A}_1P_{\rm active}\geq 0$, then $P_{\rm con}$ is a decreasing function of $r_g$,
\item If $1-\mathcal{A}_1P_{\rm active}\leq 0$, then $P_{\rm con}$ is an increasing function of $r_g$.
\end{enumerate}
Consequently, we can infer that, since $0\leq P_{\rm acitve}\leq 1$, $P_{\rm con}$ is a decreasing function of $r_g$ as long as $\mathcal{A}_1\leq 1$. In the case of $\mathcal{A}_1\geq 1$, the relation between $P_{\rm con}$ and $r_g$ can be explained as follows: (i) $P_{\rm con}$ is an increasing function of $r_g$ as long as $P_{\rm active}\geq \frac{1}{\mathcal{A}_1}$ (or $r_g\leq \sqrt{\frac{\ln(\mathcal{A}_1)}{\lambda_S\pi}}$), and (ii) $P_{\rm con}$ is a decreasing function of $r_g$ as long as $P_{\rm active}\leq \frac{1}{\mathcal{A}_1}$ (or $r_g\geq \sqrt{\frac{\ln(\mathcal{A}_1)}{\lambda_S\pi}}$).
This concludes the proof.
\section{Proof of Theorem~\ref{thm:Psec}}\label{app:Psec}
From Definition~\ref{def:2} of $P_{\rm sec}$, we observe that we need to jointly analyze the values of ${\rm SINR_S}(y_j)$ at all the locations $y_j\in\Phi_S$. Despite the usual assumption throughout most of the stochastic geometry-based literature on secrecy analysis that these values are uncorrelated, this is actually not precise. The reason for that is the dependence of ${\rm SINR_S}(y_j)$, by definition, on the PPP $\Phi_P$ for all $y_j\in\Phi_S$. Some recent works started working on characterizing the correlation between intereference levels at different locations~\cite{krishnan2016spatio}. However, most of these works focus on characterizing the correlation between only two locations assuming the knowledge of the distance between them. Unfortunately, these results will not be useful for our analysis. Hence, aligning with the existing literature, we will ignore this correlation in our analysis with the knowledge that this will provide an approximation. Furthermore, the accuracy of this approximation is expected to get worse as the value of $\lambda_S$ increases. This is due to the fact that the distances between ERs decrease as $\lambda_S$ increases, which was shown in~\cite{krishnan2016spatio} to increase the correlation. For notational simplicity, and without any loss of generality due to the stationarity of PPP, we will assume that the typical PT is placed at the origin, i.e. $x_1=o$, in the rest of this proof. All the analysis provided in this section is conditioned on the event $R_e\geq r_g$. Following the same approach as in Appendix~\ref{app:Pcon} of approximating the PHP $\Psi$ with a PPP $\Psi_P$, and defining $I_2(y_j)=\sum_{x_i\in\Psi_P\backslash x_1}g_{i,j}\|x_i-y_j\|^{-\alpha}$, $P_{\rm sec}$ can be derived as follows
\begin{align}
\label{app:6}
P_{\rm sec}&=\mathbb{E}_{\Phi_S,{I_2},\{g_{1,j}\}}\left[\mathbbm{1}\left(\bigcup_{y_j\in\Phi_S}\frac{g_{1,j}\|y_j\|^{-\alpha}}{I_2(y_j)+\frac{\sigma_S^2}{P_t}}\leq\beta_S\right)\right]\stackrel{(g)}{=}\mathbb{E}_{\Phi_S,{I_2},\{g_{1,j}\}}\left[\prod_{y_j\in\Phi_S}\mathbbm{1}\left(\frac{g_{1,j}\|y_j\|^{-\alpha}}{I_2(y_j)+\frac{\sigma_S^2}{P_t}}\leq\beta_S\right)\right]\nonumber\\
&\stackrel{(h)}{=}\mathbb{E}_{\Phi_S,{I_2}}\left[\prod_{y_j\in\Phi_S}\left(1-\exp\left(-\frac{\beta_S\left(I_2(y_j)+\frac{\sigma_S^2}{P_t}\right)}{\|y_j\|^{-\alpha}}\right)\right)\right]\nonumber\\
&\stackrel{(i)}{=}\mathbb{E}_{\Phi_S}\left[\prod_{y_j\in\Phi_S}\left(1-\exp\left(-\frac{\beta_S\left(\frac{\sigma_S^2}{P_t}\right)}{\|y_j\|^{-\alpha}}\right)\mathbb{E}\left[\exp\left(-\frac{\beta_SI_2(y_j)}{\|y_j\|^{-\alpha}}\right)\right]\right)\right],
\end{align}
where step $(g)$ (and step $(i)$) follow from assuming that the values of ${\rm SINR_S}(y_j)$ (and $I_2(y_j)$) are uncorrelated, as we discussed earlier in this Appendix. Step $(h)$ is due to assuming the set of fading gains $\{g_{1,j}\}$ to be i.i.d with $g_{1,j}\sim \exp(1)$. Defining the Laplace transform of $I_2(y_j)$ by $\mathcal{L}_{I_2(y_j)}(s)=\mathbb{E}[\exp(-sI_2(y_j))]$, we note that there is only one difference in the derivation of $\mathcal{L}_{I_2(y_j)}(s)$ compared to that of $\mathcal{L}_{I}(s)$ in Eq.~\ref{app:3}. The difference is in the reference point from where we are observing the interference. In Appendix~\ref{app:Pcon}, the reference point was the PR, which does not have a minimum distance from any active interfering PT. In the current derivation, the reference point is an ER, which has a minimum distance of $r_g$ from any active interfering PT. Hence, the derivation of $\mathcal{L}_{I_2(y_j)}(s)$ will be exactly the same as in Eq.~\ref{app:3} until step $(e)$, where the minimum distance effect will appear in the lower limit of the integral as follows
\begin{align}
\label{app:7}
\mathcal{L}_{I_2(y_j)}(s)=\exp\left(-2\pi\tilde{\lambda}_P\int_{r_g}^{\infty}\frac{sr_x^{-\alpha}}{1+sr_x^{-\alpha}}r_x{\rm d}r_x\right).
\end{align}
Note that the above expression is not a function of $y_j$, so we drop it from the notation of Laplace transform. The final expression for $\mathcal{L}_{I_2}(s)$ as provided in Theorem~\ref{thm:Psec} follows after simple mathematical manipulations. Substituting Eq.~\ref{app:7} in Eq.~\ref{app:6}, we get
\begin{align}
\label{app:8}
P_{\rm sec}&=\mathbb{E}_{\Phi_S}\left[\prod_{y_j\in\Phi_S}\left(1-\exp\left(-{\beta_S\left(\frac{\sigma_S^2}{P_t}\right)}{\|y_j\|^{\alpha}}\right)\mathcal{L}_{I_2}\left({\beta_S}{\|y_j\|^{\alpha}}\right)\right)\right]\nonumber\\
&\stackrel{(k)}{=}\exp\left(-2\pi{\lambda}_S\int_{y\in\mathbb{R}^2\cap\mathcal{B}(o,r_g)}\exp\left(-{\beta_S\left(\frac{\sigma_S^2}{P_t}\right)}{\|y\|^{\alpha}}\right)\mathcal{L}_{I_2}\left({\beta_S}{\|y\|^{\alpha}}\right){\rm d}y\right),
\end{align}
where step $(k)$ results from applying PGFL of PPP, and the integration is over $y\in\mathbb{R}^2\cap\mathcal{B}(o,r_g)$ because the analysis in this section is conditioned on the event $R_e\geq r_g$, which means that the typical PT is active. Since we assumed that the typical PT is placed at the origin in this derivation, the ball $\mathcal{B}(o,r_g)$ is clear of ERs. Converting from Cartesian to polar co-ordinates leads to the final result in Theorem~\ref{thm:Psec}.
\section{Proof of Theorem~\ref{thm:Penergy}}\label{app:Penergy}
The density of successfully powered ERs can be derived as follows
\begin{align}
\label{app:9}
P_{\rm energy}=\lambda_S\mathbb{P}\left(\eta P_t R_p^{-\alpha}w\geq E_{\rm min}\right)\stackrel{(l)}{=}\mathbb{E}_{R_p}\left[\exp\left(-\frac{E_{\rm min}R_p^{\alpha}}{\eta P_t} \right)\right],
\end{align}
where $R_p$ is the distance between the ER and its nearest active PT, and step $(l)$ is due to $w\sim\exp(1)$. The distance $R_p$ represents the contact distance of a PHP observed from a hole center. Unfortunately, the exact distribution of this distance is unknown. However, the approach of approximating the PHP $\Psi$ with a PPP $\Psi_P$ is known to provide fairly tight approximation of the contact distance distribution of PHP~\cite{WCL_2}. Given that the nearest active PT to the ER is at a distance of at least $r_g$, the distribution of $R_p$ is
\begin{align}
\label{app:10}
f_{R_p}(r_p)=2\pi\tilde{\lambda}_P\exp\left(-\pi\tilde{\lambda}_P(r_p-r_g)^2\right),\ \ \ \ r_p\geq r_g.
\end{align}
Using this distribution to compute the expectation in Eq.~\ref{app:9} leads to the final result in Theorem~\ref{thm:Penergy}.
\bibliographystyle{IEEEtran}
\bibliography{hokie-HD-updated}

\end{document}